\PassOptionsToPackage{colorlinks=true,pdfborder={0 0 0},linkcolor=violet,citecolor=blue}{hyperref}
\PassOptionsToPackage{x11names,dvipsnames,table}{xcolor}
\documentclass[runningheads,review,screen,anonymous]{llncs}
\usepackage{algorithm}
\usepackage{algpseudocodex}
\usepackage{graphicx}
\usepackage{inconsolata}

\algnewcommand\algorithmicswitch{\textbf{switch}}%
\algnewcommand\algorithmiccase{\textbf{case}}%

\makeatletter
\algdef{SE}[SWITCH]{Switch}{EndSwitch}[1]{\algpx@startIndent\algpx@startCodeCommand\algorithmicswitch\ #1\ \algorithmicdo}{\algpx@endIndent\algpx@startCodeCommand\algorithmicend\ \algorithmicswitch}%
\algdef{SE}[CASE]{Case}{EndCase}[1]{\algpx@startIndent\algpx@startCodeCommand\algorithmiccase\ #1}{\algpx@endIndent\algpx@startCodeCommand\algorithmicend\ \algorithmiccase}%

\ifbool{algpx@noEnd}{%
  \algtext*{EndSwitch}%
  \algtext*{EndCase}%
  \apptocmd{\EndSwitch}{\algpx@endIndent}{}{}%
  \apptocmd{\EndCase}{\algpx@endIndent}{}{}%
}{}%

\pretocmd{\Switch}{\algpx@endCodeCommand}{}{}
\pretocmd{\Case}{\algpx@endCodeCommand}{}{}

\ifbool{algpx@noEnd}{%
  \pretocmd{\EndSwitch}{\algpx@endCodeCommand[1]}{}{}%
  \pretocmd{\EndCase}{\algpx@endCodeCommand[1]}{}{}%
}{%
  \pretocmd{\EndSwitch}{\algpx@endCodeCommand[0]}{}{}%
  \pretocmd{\EndCase}{\algpx@endCodeCommand[0]}{}{}%
}%
\makeatother

\usepackage[
  disable,     
  textwidth=\marginparwidth,
  colorinlistoftodos,
  prependcaption,
  linecolor=green,backgroundcolor=green!25,bordercolor=green
]{todonotes}
\makeatletter
\if@todonotes@disabled
\else 
  \newlength{\increase}
  \paperwidth=\dimexpr \paperwidth + \increase\relax
  \oddsidemargin=\dimexpr\oddsidemargin + .5\increase\relax
  \evensidemargin=\dimexpr\evensidemargin + .6\increase\relax
  \marginparwidth=\dimexpr \marginparwidth + .75\increase\relax
  \let\tmptitle\title
  \renewcommand{\title}[1]{\tmptitle{#1{\tiny\normalfont \makebox[10em]{\colorbox{red!20}{todos enabled}}}}}
\fi
\makeatother

\usepackage{array}
\usepackage{amsmath}
\usepackage{mathtools}
\usepackage{amssymb}

\usepackage{amsfonts}
\usepackage{wrapfig}
\usepackage{adjustbox}
\usepackage[subtle,floats=normal]{savetrees}
\usepackage{microtype}
\usepackage{mdframed}
\usepackage{nicefrac}

\usepackage[%
  colorlinks=true,
  pdfborder={0 0 0},
  linkcolor=violet,
  citecolor=blue
]{hyperref}
\usepackage{xurl}
\usepackage[capitalise]{cleveref}

\usepackage{semantic}
\usepackage{makecell}
\usepackage{xspace}

\usepackage{pgfplots}
\usepackage{listofitems}

\usepackage{tikz}
\usetikzlibrary {
  shapes.geometric,
  arrows.meta,
  bending,
  positioning,
  calc,
  decorations.pathreplacing,
  fit
}

\usepackage{pifont}
\newcommand{\xmark}{\text{\ding{55}}}
\newcommand{\cmark}{\text{\ding{51}}}

\usepackage{listings}

\makeatletter
\lst@AddToHook{OnEmptyLine}{\addtocounter{lstnumber}{-1}}
\makeatother

\crefname{lstlisting}{listing}{listings}
\Crefname{lstlisting}{Listing}{Listings}

\newcommand{\lsem}{\left[\hspace{-.2em}\left[}
\newcommand{\rsem}{\right]\hspace{-.2em}\right]}
\newcommand{\rgra}{\right\rangle\hspace{-.2em}\right\rangle}
\newcommand{\lgra}{\left\langle\hspace{-.2em}\left\langle}

\newcommand\YAMLkeystyle{\color{black}\ttfamily}
\newcommand\YAMLvaluestyle{\color{blue}\ttfamily}

\makeatletter

\newcommand\language@yaml{yaml}

\expandafter\expandafter\expandafter\lstdefinelanguage
\expandafter{\language@yaml}
{
keywords={true,false,null,y,n,fail},
keywordstyle=\ttfamily\color{blue},
basicstyle=\YAMLkeystyle,                                 
sensitive=false,
comment=[l]{\#},
morecomment=[s]{/*}{*/},
commentstyle=\color{purple}\ttfamily,
stringstyle=\YAMLvaluestyle\ttfamily,
moredelim=[l][\color{orange}]{\&},
moredelim=[l][\color{magenta}]{*},
moredelim=**[il][\hlstr{:}\YAMLvaluestyle]{:},   
morestring=[b]',
morestring=[b]",
literate =    {---}{{\ProcessThreeDashes}}3
{>}{{\textcolor{red}\textgreater}}1
{|}{{\textcolor{red}\textbar}}1
{\ -\ }{{\mdseries\ -\ }}3,
}

\lstdefinelanguage{Lisp}{
morekeywords={goal,exists,=},
keywordstyle=\ttfamily\color{RoyalBlue},
literate={=}{{{\color{RoyalBlue}\ttfamily\bfseries =}\color{black}}}{1}
{:}{{{\color{RoyalBlue}\ttfamily\bfseries :}\color{black}}}{1}
{(}{{{\color{gray}\ttfamily\bfseries (}\color{black}}}{1}
        {)}{{{\color{gray}\ttfamily\bfseries )}\color{black}}}{1}
  }

\definecolor{base03}{RGB}{0, 43, 54}
\definecolor{base02}{RGB}{7, 54, 66}
\definecolor{base01}{RGB}{88, 110, 117}
\definecolor{base00}{RGB}{101, 123, 131}
\definecolor{base0}{RGB}{131, 148, 150}
\definecolor{base1}{RGB}{147, 161, 161}
\definecolor{base2}{RGB}{238, 232, 213}
\definecolor{base3}{RGB}{253, 246, 227}
\definecolor{yellow}{RGB}{181, 137, 0}
\definecolor{orange}{RGB}{203, 75, 22}
\definecolor{red}{RGB}{220, 50, 47}
\definecolor{magenta}{RGB}{211, 54, 130}
\definecolor{violet}{RGB}{108, 113, 196}
\definecolor{blue}{RGB}{38, 139, 210}
\definecolor{cyan}{RGB}{42, 161, 152}
\definecolor{green}{RGB}{133, 153, 0}

\lstdefinelanguage{Python}{
  keywords={typeof, null, catch, switch, in, int, str, float, self, throw, import, return, class, if, elif, endif, while, do, else, for, catch, def, raise, not},
  keywordstyle=\color{violet}\ttfamily,
  ndkeywords={True, False},
  ndkeywordstyle=\color{red}\ttfamily,
  identifierstyle=\color{black}\ttfamily,
  sensitive=false,
  comment=[l]{\#},
  morecomment=[s]{/*}{*/},
  commentstyle=\color{base00}\ttfamily,
  stringstyle=\color{green}\ttfamily,
  morestring=[b]',
  morestring=[b]"
}

\newenvironment{codeEnv}[1][base3]
{\begin{mdframed}[backgroundcolor=#1,innertopmargin=-2,innerbottommargin=-2,linecolor=#1!80!black]}
    {\end{mdframed}}

\newcommand{\code}[1]{\lstinline[language=yaml]{#1}}
\newcommand{\mcode}[1]{\mbox{\lstinline[language=yaml]{#1}}}
\lst@AddToHook{EveryLine}{\ifx\lst@language\language@yaml\YAMLkeystyle\fi}
\makeatother

\newcommand\ProcessThreeDashes{\llap{\color{cyan}\mdseries-{-}-}}

\usepackage{tikz}

\definecolor{E}{RGB}{51,153,255}

\definecolor{S}{RGB}{6,153,0}

\definecolor{B}{RGB}{153,0,0}

\newcommand{\hln}[1]{{\color{RoyalBlue}\texttt{#1}}}
\newcommand{\hlopt}[1]{{\color{purple}\texttt{#1}}}
\newcommand{\hlstr}[1]{{\color{black}\textbf{\texttt{#1}}}}
\newcommand{\many}[1]{\overline{#1}}
\newcommand{\Div}{\ |\ }
\DeclareMathOperator{\dom}{dom}

\newcommand{\Tags}{\mathcal{T}}
\newcommand{\App}{\mathcal{P}}
\newcommand{\Policies}{\App}
\newcommand{\policy}{p}
\newcommand{\Blocks}{\mathcal{B}}
\newcommand{\block}{b}
\newcommand{\Funcs}{\mathcal{F}}

\newcommand{\Works}{\mathcal{W}}
\newcommand{\rName}[1]{{\small{\([\mathit{#1}]\)}}}
\newcommand{\emptyseq}{\varepsilon}
\newcommand{\auxfn}[1]{\textbf{\textsf{#1}}}
\newcommand{\strat}{\auxfn{strategy}}
\newcommand{\valid}{\auxfn{valid}}

\newcommand{\fopt}{\textit{f\_opt}}
\newcommand{\wopt}{\textit{w\_opt}}
\newcommand{\sopt}{\textit{s\_opt}}
\newcommand{\iopt}{\textit{i\_opt}}
\newcommand{\aopt}{\textit{a\_opt}}
\newcommand{\enc}[1]{{\color{green}\lsem{\color{black}#1}\rsem}}
\newcommand{\encp}[1]{{\color{purple}\lgra{\color{black}#1}\rgra}}

\newcommand{\reg}{\textit{reg}}

\newcommand{\mOpt}[1]{\colorbox{gray!12}{\(#1\)}}

\newcommand{\Reach}{\texttt{Reach}}
\newcommand{\CoOccur}{\texttt{CoOccur}}
\newcommand{\simple}{\texttt{simple}}
\newcommand{\filter}{\texttt{fltr}}

\newcommand{\app}{\textsf{APP}\xspace}
\newcommand{\appp}{\texorpdfstring{\textsf{\MakeLowercase{a}APP}}{}\xspace}
\newcommand{\rulelabel}[1]{\rule[.2em]{5em}{.1px} \quad \textnormal{\color{gray}#1} \quad \rule[.2em]{5em}{.1px}}

\Crefname{equation}{Eq.}{Eqs.}
\Crefname{figure}{Fig.}{Figs.}
\Crefname{tabular}{Tab.}{Tabs.}
\Crefname{section}{Sec.}{Secs.}

\AtBeginDocument{%
  }

\author{
Giuseppe De Palma\(^{1,2}\) \and 
Saverio Giallorenzo\(^{1,2}\) \and
\\Jacopo Mauro\(^{3} \)\and
Matteo Trentin\(^{1,2,3}\) \and
Gianluigi Zavattaro\(^{1,2}\)
}

\institute{
  Alma Mater Studiorum - Universit\`a di Bologna, Italy
  \and OLAS team INRIA, France
  \and University of Southern Denmark, Denmark}

\begin{document}

\title{On the Complexity of Reachability Properties in Serverless Function Scheduling}

\maketitle

\begin{abstract}
Functions-as-a-Service (FaaS) is a Serverless Cloud paradigm where a platform
manages the execution scheduling (e.g., resource allocation, runtime
environments) of stateless functions. Recent developments demonstrate the
benefits of using domain-specific languages to express per-function scheduling
policies, e.g., enforcing the allocation of functions on nodes that
enjoy low data-access latencies thanks to proximity and connection pooling.

We present \appp, an affinity-aware extension of a platform-agnostic function
scheduling language. We formalise its scheduling semantics
and then study the complexity of statically checking reachability properties,
e.g., useful to verify
that trusted and untrusted functions cannot be co-located. Analysing different fragments of \appp, we show that checking reachability of policies without affinity has linear complexity, while affinity makes the problem PSpace.


\end{abstract}


\section{Introduction}
\label{sec:intro}

Functions-as-a-Service (FaaS) is a programming paradigm supported by the Serverless Cloud
execution model~\cite{Jonas-etal:BerkeleyViewOnServerless}. In FaaS,
developers implement a distributed architecture by composing
stateless functions and delegate concerns like execution runtimes and resource allocation to the serverless platform, thus focusing on writing code that implements business logic rather than worrying about infrastructure management.
The main cloud
providers offer FaaS~\cite{web:IntroducingAwsLambda,web:googlefunctions,web:azurefunctions} and open-source alternatives exist too~\cite{web:OpenWhisk,web:openfaas,HSHVAA16,web:fission}.

A common denominator of these platforms is that they manage the allocation of
functions over the available computing resources, also called \emph{workers},
following opinionated policies that favour some performance principle.
Indeed, effects like \emph{code locality}~\cite{HSHVAA16}---due to latencies in
loading function code and runtimes---or \emph{session
locality}~\cite{HSHVAA16}---due to the need to authenticate and open new
sessions to interact with other services---can substantially increase the run time of
functions.
The breadth of the design space of serverless scheduling policies is
witnessed by the growing literature focused on techniques that mix one or more
of these locality principles to increase the performance of function
execution, assuming some locality-bound traits of functions~\cite{SSGP17,OYZHHAA18,KNJ18,ABE18,SBW19,SG19,AHKSNV19,KSB19,SA20,SP20,SFP20,SWLSGHT20,SFGCBCLTRB20,CAVJLLPPR20,KGB20,BS21,BQ21,ZE21,SAVA21,SJCGB22}.
Besides performance, functions can have functional requirements that the scheduler shall consider. For example, users might want to ward off
allocating their functions alongside ``untrusted'' ones---common threat vectors
in serverless are limited function isolation and the ability of functions to
(surreptitiously) gather weaponisable information on the runtime, the
infrastructure, and the other
tenants~\cite{BCCCFIMMRS17,WLZRS18,AFFRSSW18,PPTMAA20}.

Although one can mix different principles to expand the profile coverage of a
given platform-wide scheduler policy, the latter hardly suits all kinds of
scenarios. This shortcoming motivated the introduction of a domain-specific,
platform-agnostic, declarative language, called \emph{Allocation Priority
Policies} (\app), for specifying custom function allocation
policies~\cite{PGMZ20,DGMTZ22}.
Thanks to \app, the same platform can support different scheduling policies,
each tailored to meet the specific needs of a set of related functions.
%
%
\app has been validated by implementing a serverless platform as an extension of
the open-source Apache OpenWhisk project---the \app-based variant outperforms
vanilla OpenWhisk in several locality-bound scenarios~\cite{PGMZ20,DGMTZ22}.

Our contributions originate from the observation of lower levels of a typical
cloud application stack, where popular Infrastructure-as-a-Service (IaaS)
platforms (e.g., OpenStack~\cite{web:openstack_affinity}), and
Container-as-a-Service (CaaS) systems (e.g.,
Kubernetes~\cite{web:kube_affinity}) allow users to express affinity and
anti-affinity constraints about the
allocation of VM/containers---e.g., to reduce overhead by shortening data paths
via co-location, to increase reliability by evenly distributing VM/containers
among different nodes, and for security, such as preventing the co-location of
VM/containers belonging to different trust tiers. 
On the contrary, FaaS platforms do not natively support the possibility to
express affinity-aware scheduling,
where function allocation depends on the presence (affinity) or absence
(anti-affinity) at scheduling time of other functions in execution on the
available workers.
%
Since, serverless applications can also benefit from affinity-aware scheduling,
starting from \app, which works only at FaaS level and is platform-agnostic, we
study the addition of affinity and anti-affinity constraints at the FaaS level
by proposing a
new \emph{affinity-aware} extension, called \appp---presented in \cref{sec:app}.

\emph{Contributions}
From the formal methods point of view, we 
focus on two related analysis problems for \appp systems: \emph{reachability}
(i.e., checking if a function $f$ can be deployed on a targeted worker) and
\emph{co-occurrence} (i.e., checking if two function $f$ and $g$  can be
colocated, so they can run simultaneously on a targeted worker).
The motivation behind this study is twofold. On the one hand, the analysis of
reachability and co-occurrence is important to verify the correctness of \appp
scripts. In fact, \appp scripts are used to express function scheduling policies
and their correctness can be verified by replying to questions like: can this
critical function $f$ be scheduled on an untrusted worker $w$? Can this
trustworthy function $f$ be scheduled on a worker at the same time of another
suspicious function $g$? Moreover, the investigation of these problems, in
particular of their complexity, sheds light on the expressiveness of \appp,
intended as the possibility to express complex scheduling policies for which
reachability and co-occurrence problems are hard to verify. In this paper, we
mainly focus on this second line of research (an initial investigation
of the implementation and exploitation of reachability and co-occurrence 
analysers can be found in \cite{PalmaGMTZ23}).

We start by investigating the complexity of reachability and co-occurrence in
\app, i.e., the fragment of \appp without (anti-)affinity constraints, and we
prove that both problems have linear-time complexity
(\cref{sec:app_properties}). Then, we show that the problems become
PSpace-complete in \appp (\cref{sec:appp}): this result formally witnesses that
the \appp extension allows for the specification of much harder systems to
verify. We solidify this result by showing that the jump in complexity comes
from affinity constraints since, without them, \emph{reachability} and
\emph{co-occurrence} preserve a linear complexity while with affinity only
(i.e., without anti-affinity) the problems become NP-hard
(\cref{sec:appp_other}).

\section{\appp language}
\label{sec:app}

\begin{figure}[t]
\begin{minipage}{.66\textwidth}
\begin{adjustbox}{width=.9\textwidth}
\begin{minipage}{\textwidth}
\vspace{-1.5em}
\[
\begin{array}{lll}
&& \hspace{.12\textwidth} \textit{id} \in \textit{Identifiers} 
\hspace{.1\textwidth} n \ \in \ \mathbb{N} 
\\[.5em]
\textit{app}         & \Coloneqq & \many{- \textit{tag}}
\\
\textit{tag}         & \Coloneqq & \textit{id} \ \hlstr{:} \ \many{\mathtt{-}\ \textit{block} } \quad \mOpt{\hln{followup}\ \hlstr{:}\ \fopt}
\\
\textit{block}       & \Coloneqq & \hln{workers} \ \hlstr{:}\ \wopt
\quad \mOpt{\hln{strategy} \ \hlstr{:}\ \sopt}\\&&
\quad \mOpt{\hln{invalidate} \ \hlstr{:}\ \many{-\ \iopt}}
\quad \mOpt{\hln{affinity} \ \hlstr{:}\ \many{-\ \aopt}}
\\
\wopt      & \Coloneqq & \hlopt{*} \Div \many{\mathtt{-}\ \textit{id}}
\\
\sopt      & \Coloneqq & \hlopt{any} 
              \Div \hlopt{best\_first}
\\
\iopt      & \Coloneqq & \hlopt{capacity\_used} \ n \hlopt{\%} \Div \hlopt{max\_concurrent\_invocations} \ n 
\\
\aopt & \Coloneqq & \textit{id} \Div \textit{!id}
\\
\fopt      & \Coloneqq & \hlopt{default} \Div \hlopt{fail}
\end{array}
\]
\end{minipage}
\end{adjustbox}
\caption{\label{fig:app_syntax} \appp syntax.}
\end{minipage}
\hspace{-1.5em}
{\color{gray}\vrule}\hspace{0em}
\begin{minipage}{.35\textwidth}
\begin{adjustbox}{width=\textwidth}
\begin{minipage}{1.18\textwidth}
\vspace{.5em}
\begin{codeEnv}
\begin{lstlisting}[xleftmargin=-1.2em, language=yaml, mathescape, basicstyle=\linespread{1}\footnotesize\ttfamily,numbers=none]
- f_tag:
  - $\hln{workers}$:
      - local_w1
      - local_w2
    $\hln{strategy}\hlstr{:}\ \hlopt{best\_first}$
    $\hln{invalidate}$:
      - $\hlopt{capacity\_used}$ 80%
    $\hln{affinity}\hlstr{:}$ g_tag,!h_tag
  - $\hln{workers}$:
      - public_w1
  $\hln{followup}\hlstr{:}\ \hlopt{fail}$
\end{lstlisting}
\end{codeEnv}
\end{minipage}
\end{adjustbox}
\caption{\label{fig:app_example} Example \appp script.}
\end{minipage}
\vspace{-2em}
\end{figure}

In this section, we present \appp, our extension of the
FaaS function scheduling language \app~\cite{PGMZ20,DGMTZ22} 
with affinity and anti-affinity constraints. 

We report in \cref{fig:app_syntax} the syntax of \appp. From here on, we
indicate syntactic units in \emph{italics}, optional fragments in
\(\mOpt{grey}\), terminals in \code{monospace}, and lists with \(\many{bars}\).
The syntax of \appp draws inspiration from YAML~\cite{web:YAML}, a renowned
data-serialisation language for configuration files---e.g., many modern tools
use the format, like Kubernetes, Ansible, and Docker.\footnote{While \appp
scripts are YAML-compliant, for presentation, we slightly stylise the syntax to
increase readability. For instance, we omit quotes around strings, e.g.,
\hlopt{*} instead of \hlopt{"*"}.}
The idea behind \appp is that functions have associated a tag that identifies
some scheduling policies. An \appp script represents: \textit{i)} named
scheduling policies identified by a \textit{tag} and \textit{ii)} policy
\(\textit{block}\)s that indicate either some collection of workers, each
identified by a worker \textit{id}, or the universal \hlopt{*}. To schedule a
function, we use its tag to retrieve the scheduling policy that includes one
or more blocks of possible workers. To select the worker, we iterate
top-to-bottom on the blocks. We stop at the first block that has a non-empty
list of valid workers and then select one of those workers according to the
strategy defined by the block (described later).

Each tag can define a \hln{followup} clause, which specifies what to do if the
policy of the tag did not lead to the scheduling of the function; either
\hlopt{fail}, to terminate the scheduling, or \hlopt{default} to apply the
special \texttt{default}-tagged policy.
Each block can define a \hln{strategy} for worker selection ($\hlopt{any}$
selects non-deterministically one of the available workers in the list;
$\hlopt{best\_first}$ selects the first available worker in the list), a list of
constraints that \hln{invalidate}s a worker for the allocation
($\hlopt{capacity\_used}$ invalidates a worker if its resource occupation
reaches the set threshold; $\hlopt{max\_concurrent\_invocations}$ invalidates a
worker if it hosts more than the specified number of functions), and an
\hln{affinity} clause that carries a list containing affine tag identifiers
$\textit{id}$ and anti-affine tags, represented by negated tag identifiers
$\textit{!id}$. \appp is a minimal extension of \app adding the possibility to
use this latter \hln{affinity} construct that is not available in the original
\app proposal.

As an example, \cref{fig:app_example} show an \appp policy for functions tagged
\texttt{f\_tag}. The policy has two blocks. The former restricts the allocation
of the function on the workers labelled \lstinline[language=yaml]{local_w1} and
\lstinline[language=yaml]{local_w2} and the latter on
\lstinline[language=yaml]{public_w1}. The first block specifies as invalid
(i.e., which cannot host the function under scheduling) the workers that reach a
memory consumption above 80\%. Since the strategy is $\hlopt{best\_first}$, we
allocate the function on the first valid worker; if none are valid, we proceed
with the next block. The function has affinity with $g\_tag$ and anti-affinity
with $h\_tag$. Hence, a valid worker requires the presence of at least a
function with tag $g\_tag$ and no functions with tag $h\_tag$. 
If both the first and second blocks did not find a valid worker, the scheduling
of the function \hlopt{fail}, without trying other policies.

%


The semantic of \appp is defined by an LTS over the possible configurations of
the FaaS platform.
We report a fully formalized definition of the LTS in \cref{appendix-semantics} and present a more compact and readable definition of the logic of \appp's function scheduling algorithm
as pseudo-code.
The transitions in the LTS are in the form
$C \xRightarrow{\lambda}C'$
where $C$ and $C'$ are two configurations and $\lambda$ is
a label describing the performed action. There are three
kinds of labels: i) $(start,f,w)$
indicates the allocation of an instance of the function $f$ on the worker $w$,
ii) $(done,f,w)$ denotes the deallocation on $f$ on $w$, and iii)
$(\mathit{fail},f)$ traces the failure to schedule \(f\) on the current
configuration.

We now specify the domains and structures used in the semantics.
We use $\Works$, ranged over by $w$, to denote the set of
workers, while $\Funcs$, ranged over by $f$, denotes the set of functions. We
use $\mathcal C$, ranged over by $C$, to denote the set of platform
configurations. A configuration associates each of its workers (in \(\Works\))
with a triple relating the multiset of functions (\(\mathit{Multiset}(\Funcs)\))
currently allocated on that worker, the amount of resources (in \(\mathbb{N}\))
used by such functions, and the maximal amount of resources (also in
\(\mathbb{N}\)) available to that worker. Functions are tagged to associate them
with a scheduling policy.
We use $\Tags$, ranged over by $t$, to denote the set of tags and define \reg{}
(short for registry) as a map that associates each function with its tag and its
occupancy, i.e., the amount of resources needed to host it. \(\mathbb{N}\)
represents the natural numbers---even considering fractional resources, we deem
naturals enough fine-grained for our purpose, since we can always convert these
to \(\mathbb{N}\) with a constant multiplying factor.

We use (Python-like) pseudo-code in \cref{lst:code} to present \appp's
scheduling logic.

\begin{figure}[t]
\noindent\begin{adjustbox}{width=1.1\textwidth}
\begin{minipage}{1.4\textwidth}
\begin{lstlisting}[language=Python,numberblanklines=false,label=lst:code,caption={The pseudo-code of the \code{schedule} and \code{valid} functions.}]
def schedule(f, conf, aapp, reg):
  (memory, tag) = reg[f]
  blocks = aapp[tag].blocks # get the blocks
  if aapp[tag].followup != 'fail':
    blocks += aapp['default'].blocks # add default tag blocks
  for block in blocks:
    if '*' in block['workers']: 
      block['workers'] = conf.keys
    workers = [ for worker in block['workers'] if valid(f,worker,conf,reg,block)]
    if len(workers) > 0: # if at least one valid worker is found
      if block['strategy'] == 'best_first': 
        return workers[0]
      elif block['strategy'] == 'any': 
        return random.choice(workers)
  raise Exception('Function not schedulable')

def valid(f, w, conf, reg, block):
  (memory, tag) = reg[f]
  if (w not in conf) or (conf[w]['memory_used'] + memory > conf[w]['max_memory']):
    return False
  if 'invalidate' in block:
    if ('capacity_used' in block['invalidate']) and
          (block['invalidate']['capacity_used'] <= conf[w]['memory_used']):
        return False
    if ('max_concurrent_invocations' in block['invalidate']) and
        (block['invalidate']['max_concurrent_invocations'] <= len(conf[w]['fs'])):
      return False
  if 'affinity' in block:
    affine_tags = set([t for t in block['affinity'] if not t.startswith('!')])
    anti_affine_tags = set([t[1:] for t in block['affinity'] if t.startswith('!')])
    w_tags = set([t for (_, t) in [reg(f) for f in conf[w]['fs']]])
    for t in affine_tags:
      if t not in w_tags: return False
    for t in anti_affine_tags:
      if t in w_tags: return False
  return True
\end{lstlisting}
\end{minipage}
\end{adjustbox}
\end{figure}


The function \lstinline[language=yaml]{schedule} requires the name of the function to be scheduled (\lstinline[language=yaml]{f}), the map representing the configuration (\lstinline[language=yaml]{conf}), the \appp script encoded as a Python dictionary of objects (\lstinline[language=yaml]{aapp}), and the registry mapping the memory occupation and the tag for every function (\lstinline[language=yaml]{reg}).
For readability, we treat the tuple representing the configuration of a worker as a map, denoting with \lstinline[language=yaml]{fs}, \lstinline[language=yaml]{memory_used}, and \lstinline[language=yaml]{max_memory} respectively the list of functions already scheduled on the node, the memory allocated for those functions and the total amount of memory of the worker.

The function \lstinline[language=yaml]{schedule} first gets the tag associated with $f$ (Line 2) and then extracts the blocks associated with this tag in the \lstinline[language=yaml]{appp} script (Line 3). If the follow-up strategy is different from ``fail'' the blocks associated with the \lstinline[language=yaml]{default} tag are appended to the list of $f$'s bocks (Line 5). Then for every block in order of appearance, the list of valid workers is obtained (Line 9).
Note that if the \lstinline[language=yaml]{*}  is used, then all the workers present in the configuration are checked (Line 8).
If the list of valid workers is nonempty, the first one is chosen if the strategy is \lstinline[language=yaml]{best_first} (Line 12), a random one otherwise (Line 14). The schedule fails if no blocks have valid workers (Line 15).

The validity of the workers is computed by the function \lstinline[language=yaml]{valid} that ensures that the worker is available and has enough capacity to host the function (Lines 18-19), that the capacity\_used and max\_concurrent\_invocations are satisfied (Lines 21-26), and that the (anti-)affinity constraints, if any, are satisfied (lines 27-34).

The three kinds of transitions in the LTS are defined as follows.
In case \lstinline[language=yaml]{schedule(f,C,aapp,reg)} returns \lstinline[language=yaml]{w} then $C \xRightarrow{(start,f,w)}C'$ where $C'$ is the configuration $C$ in which the worker \lstinline[language=yaml]{w} has allocated an additional function \lstinline[language=yaml]{f}. If the schedule invocation fails then $C \xRightarrow{(\mathit{fail},f)}C$. Finally, 
the termination of the execution of a function is represented as a non-deterministic transition $C \xRightarrow{(done,f,w)}C'$ that is always possible provided that in the configuration $C$ a function \lstinline[language=yaml]{f} is scheduled on worker \lstinline[language=yaml]{w}.
Here  $C'$ is the configuration $C$ in which an instance of function
\lstinline[language=yaml]{f} is deallocated from the worker
\lstinline[language=yaml]{w}.

As a final remark, we note that in \appp the relation of (anti-)affinity is ``directional''---similarly to the
one introduced by Microsoft in its IaaS
offering~\cite{service_fabric_affinity}. In particular, we do not impose any properties
like symmetry 
on affinity or anti-affinity. One might argue
that imposing 
properties or well-formedness guarantees can
prevent programmers from making mistakes in their \appp{} scripts (e.g., they
can misconfigure the policies of two functions that they wanted to be mutually
anti-affine because they forgot to include a constraint in some block). While
avoiding these occurrences is important, our objective in this paper is to allow
\appp{} to capture as many useful scenarios as possible and imposing
well-formedness properties would go against that goal.
\footnote{As an example, if we had
  symmetric anti-affinity, then we would not capture a scenario in which a
  function \code{init} is the seeding function for a database and function
  \code{query} manipulates that data. Function \code{init} should always run
  before \code{query} but never where \code{query} is already running, while
  function \code{query} should run where \code{init} is present. To obtain this
  behaviour, we need \code{init} anti-affine with \code{query} but \code{query}
  affine with \code{init}.}

\section{Reachability and Co-Occurrence in \app}
\label{sec:app_properties}

In this section, we formally define the \emph{reachability} 
and \emph{co-occurrence} problem and we start the investigation 
of their complexity starting with the fragment of \appp without
(anti-)affinity constraints (i.e., we consider only the \app language).

Before proceedings with the formal definitions, we
recap the notation we use for denoting workers $\Works$,
functions $\Funcs$, configurations $\mathcal C$
(mapping workers to triples composed of
 the multiset of the currently scheduled functions, 
 the currently occupied memory and the worker's total memory),
function tags $\Tags$,
and registry {\reg} 
(mapping functions to pairs composed of their memory occupancy
 and their tag).
An \appp policy $\policy$  is formally defined as a map from tags to policy blocks and a policy block $\Blocks$ is a tuple defining the workers of the block, the strategy, the invalidate conditions, and the lists of the affinity and anti-affinity constraints.
\begin{align*}
  w \in \Works \subset \mathit{IDs}
   &  &
  \hspace{8em}f \in \Funcs
   &  &
  \hspace{3em}C \in \mathcal C \triangleq \Works \to \mathit{Multiset}(\Funcs) \times \mathbb{N} \times \mathbb{N}
  \\
  t \in \Tags \subset \mathit{IDs}
   &  &
  \reg \in \Funcs \to \mathbb{N} \times \Tags
   &  &
  \\
  \multicolumn{6}{c}{\(
    \policy \in \Policies \triangleq \Tags \to \mathit{List}(\Blocks)
    \quad
    \Blocks \triangleq (\mathit{List}(\Works) \cup \hlopt{*}) \times \textit{\sopt} \times \mathit{List}(\textit{\iopt}) \times \mathit{List}(\mathit{IDs}) \times \mathit{List}(\mathit{IDs})\)}
\end{align*}

For simplicity and w.l.o.g., in the remainder of this section,
we assume each function is associated with its namesake tag, i.e., \(\reg(f) =
(\cdot,f)\), and refer to ``the blocks corresponding to the function'' instead
of the longer ``the blocks of the (namesake) tag corresponding to the
function''. We use the notation \(C
\xhookrightarrow{\lambda_1::\cdots::\lambda_n}C'\) as a shorthand for the
(possibly empty) sequence of LTS transitions \(C \xRightarrow{\lambda_1} \cdots
\xRightarrow{\lambda_n} C'\).


\begin{definition}[Reachability]
  Given a policy $p$, a function registry \(reg\), a configuration $C$, a
  function $f$, and a worker $w$ s.t. $C(w)=(\sigma,\cdot,\cdot)$ and $f \not\in
    \sigma$, the {\em reachability} problem \Reach($p,reg,C,f,w$) consists
  of checking if there exists $\many{\lambda}$ such that $C
    \xhookrightarrow{\many{\lambda}}C'$, assuming policy \(p\) and registry \(reg\),
  with $C'(w)=(\sigma',\cdot,\cdot)$ s.t. $f \in \sigma'$.
\end{definition}

\begin{definition}[Co-occurrence]
  Given a policy $p$, function registry \(\reg\), a configuration $C$,
  two functions $f$ and $g$, and a worker $w$ s.t. $C(w)=(\sigma,\cdot,\cdot)$ and
  $\{f,g\} \cap \sigma = \emptyset$, the {\em co-occurrence} problem \CoOccur($p,
  reg, C,\{f,g\},w$) consists of checking whether there exists $\many{\lambda}$
  such that $C \xhookrightarrow{\many{\lambda}}C'$, assuming policy \(p\) and registry \(reg\), with
  $C'(w)=(\sigma',\cdot,\cdot)$ s.t. $\{f,g\} \subseteq \sigma'$.
\end{definition}






To prove our results, we first prove a lemma that states that,
when checking the reachability of a function, we can disregard all the workers
in a given policy that are not the targeted worker.

The lemma uses the auxiliary function \(\simple\) which, given a policy
\(p\) and a worker \(w\), returns a new policy obtained by preserving only those
blocks that involve \(w\) (including the universal \(\hlopt{*}\)) and removing
all other workers therein.


Formally, \(\simple(p,w) = \{ (\ t,\ \filter(p(t), w)\ )\ |\ t \in \Tags\ \}\) where
\(\filter(\emptyseq,w) = \emptyseq\) and
 \\\(\filter((W,s,\many{i})::\many{b},w) = \begin{cases}
   (w,s,\many{i})::\filter(\many{b},w) & \mbox{if}\ W = \hlopt{*} \ \vee\ (W = \many{w} \ \wedge\ w \in \many{w})
   \\
   \filter(\many{b},w) & \mbox{otherwise}
\end{cases}\)

%



\begin{lemma}
  \label{lemma:only_target}
  Let \(C\) be an \emph{empty} configuration, i.e., \(\forall\ w \in \dom(C)\ .\
  C(w)=(\sigma,\cdot,\cdot)\ \wedge\ |\sigma| = 0\). Let $p$ be a scheduling
  policy, \(\reg\) a function registry, $f$ a function, and $w$ a worker s.t. $w
    \in \dom(C)$ and \(\exists\ \block \in \{(\many{w},\cdot,\cdot),
  (\hlopt{*},\cdot,\cdot)\} \ .\ \block \in p(f)\ \wedge\ w \in \many{w}\).

  \noindent
  We have that \Reach($p,\reg,C,f,w$) iff
  \Reach($\simple(p,w),\reg,C,f,w$).
\end{lemma}
\vspace{-4mm}

\begin{proof}
  We start from the left-to-right implication, whereby \Reach($p,\reg,C,f,w$)
  holds. Given that the configuration \(C\) is empty, \Reach($p,\reg,C,f,w$)
  implies the existence of a trace $\many{\lambda}=\lambda_1 :: \dots :: (start,
    f, w)$ such that $C \xhookrightarrow{\many{\lambda}}C'$. From \(\many{\lambda}\)
  we can obtain $\many{\lambda'}$ by removing all the labels that do not regard
  $w$, i.e., we keep only the ordered transitions in \(\many{\lambda}\) that
  either allocate or remove
  functions on $w$.

  The trace $\many{\lambda'}$ is valid for the policy $p' = \simple(p,w)$. First,
  the transitions in \(\many{\lambda'}\) regard blocks in \(p\) that contain
  \(w\), which are also those and only blocks present in
  $p'$---\(\many{\lambda'}\) also excludes possible failed scheduling attempts of
  the shape \((\textit{fail},\ g)\) from \(\many{\lambda}\), however we can safely
  remove them since these produce no change in the configuration. Second, the
  actions in \(\many{\lambda'}\) are valid also for \(p'\) since the blocks
  concerning \(w\) are the same between \(p\) and \(p'\), except for their lists
  of workers, i.e., pairwise, the two blocks have the same strategies and
  invalidation policies, except the block in \(p'\) has only \(w\) as worker.

  By definition, $\many{\lambda'}$ includes also the label $(start, f, w)$ and
  thus we have that \break\Reach($\simple(p,w),\reg,C,f,w$) holds.

  We now move to the right-to-left implication. In this case we assume that 
  \break \Reach($\simple(p,w),\reg,C,f,w$) holds, hence we know that by considering
  the policy $\simple(p,w)$ there exists a trace
  $\many{\lambda}=\lambda_1 :: \dots :: (start, f, w)$ such that $C
    \xhookrightarrow{\many{\lambda}}C'$ where \(f\) is allocated on \(w\) in \(C'\).

  Given that the transitions $(\mathit{fail},f)$ do not
  change the configuration \(C\), we can consider, w.l.o.g.,
  that $\many{\lambda}$ contains none of those labels.

  To prove this case, we proceed by induction on the length of $\many{\lambda}$
  and demonstrate that, if $C \xhookrightarrow{\many{\lambda}}C'$, with policy
  $\simple(p,w)$, then also $C \xhookrightarrow{\many{\lambda'}}C'$, with
  policy $p$, for a sequence $\many{\lambda'}$ obtained from $\many{\lambda}$ by
  adding some labels that do not concern \(w\).
  %

  The base case
  trivially holds considering an empty sequence $\many{\lambda'}$.


  We now move to the inductive case. Consider $C
    \xhookrightarrow{\many{\lambda}}C''\xRightarrow{\lambda_n} C'$. By the inductive
  hypothesis, we have that $C \xhookrightarrow{\many{\lambda'}}C''$ also holds for
  the policy $p$. We now prove that there exists a sequence ${\many{\lambda''}}$,
  that includes $\lambda_n$, such that $C''\xRightarrow{\many{\lambda''}} C'$
  assuming policy $p$.

  If $\lambda_n = (done, f, w)$, we have that $C''\xRightarrow{\lambda_n} C'$ also
  for policy $p$ because the removal of allocated functions
  occurs independently of the policy.

  If $\lambda_n = (start, f, w)$, we show that we can find a block in \(p\) that
  can perform the same action as \(\simple(p,w)\). Let \(p' =
  \simple(p,w)\) and let $b'$  
  be the block in $p'(f)$ concerned
  in \(\lambda_n\), and let $b$ be the block in $p(f)$ corresponding to $b'$ 
  (i.e., it is the block in \(p\) that originated \(b'\) in
  $\simple(p,w)$). Let $p(f)$ be the sequence of blocks $b_1, \cdots, b_k, b, \cdots$. We can
  consider that no block $b_i$, $i \in [1,k]$, can schedule the function on $w$.
  If that were not the case, we can take the \(b_i\) that allows the scheduling of
  \(f\) in \(w\) as the originating block of \(b'\) and continue our argument
  until no preceding block concerning \(w\) exists in \(p\).

  We now show that it is possible to use the block $b$ to schedule $f$ on $w$.
  This is possible only if all the blocks $b_1, \dots, b_k$ become invalid. We do
  this, in $p$, by scheduling the function $f$ on the workers of the blocks $b_1,
    \dots, b_k$ until we meet their \hln{invalidate} conditions---we reached the $\hlopt{capacity\_used}$ or $\hlopt{max\_concurrent\_invocations}$
  defined in the block or 100\% of $\hlopt{capacity\_used}$ of the
  worker (as checked by the \lstinline[language=yaml]{valid} function). When all the
  previous blocks become invalid, block $b$ becomes valid due to the inductive hypothesis (the configuration $C''$ is the same after trace $\lambda$ and
  $\lambda'$).
  We have two cases:
  \begin{itemize}
    \item
          If $b$ has strategy $\hlopt{any}$, 
          we can execute a transition labelled $(start, f, w)$, since we can select any of
          the workers of the block, including $w$.
          After this transition, we can delete the functions scheduled on all the workers
          different from $w$.
          This leads to the
          configuration $C'$ since the sequence of the mentioned transitions---the ones
          used to invalidate the blocks $b_1, \dots, b_k$, the one to allocate \(f\) on
          \(w\) \((start,f,w)\), and the removal of all functions allocated to the workers
          different from $w$---has the cumulative effect of adding one instance of $f$
          on $w$.
          %
    \item
          If $b$ has strategy $\hlopt{best\_first}$, we need to also invalidate all the
          workers that precede $w$ in the list of the block. We follow the same reasoning
          for the invalidation of the blocks that precede \(b\): we allocate $f$ on
          those workers until they are not valid anymore. After this
          additional sequence of labels, we can finally trigger $(start, f, w)$, followed
          by the removal (\rName{C_{done}}) of the functions allocated on the workers
          different from \(w\). Similarly to the previous item, this leads to the configuration $C'$.
          \qed
  \end{itemize}
\end{proof}

\noindent
We are now ready to assess the complexity of the reachability problem.


\begin{theorem}
  \label{theorem:poly_reach}
  \Reach($p,\reg,C,f,w$) has linear time complexity.
\end{theorem}
\vspace{-5mm}

\begin{proof}
  Let us assume a scheduling policy $p$. Consider the problem
  \Reach($p,\reg,C,f,w$) for a configuration $C$, a function registry \reg, a
  function $f$, and a worker $w$. Since we can always remove functions,
  we can assume, w.l.o.g., that the
  configuration $C$ is empty, i.e., with no allocated function on any of its
  workers.

  Thanks to \Cref{lemma:only_target}, we can reduce \Reach($p,\reg,C,f,w$) to
  the same reachability problem by assuming a simplified policy $\simple(p,w)$
  containing blocks where $w$ is the only present worker.
  It is sufficient to check whether one of these blocks is already valid in the
  empty configuration. In fact, if none of the blocks is valid, then
  \Reach($p,\reg,C,f,w$) does not hold since, under the current configuration,
  we cannot make an invalid block valid by adding other functions.
  This property holds because the constraints in {\app} are
    \emph{anti-monotonic}, i.e., if a worker satisfies the constraints, the same
    worker with a smaller load satisfies those constraints too.\footnote{This
    anti-monotonic property does not hold in \appp, the affinity-aware version
    of \app, where a worker with fewer functions can become invalid due to
    affinity constraints.} If, on the other hand, there exists a valid block,
    then \Reach($p,\reg,C,f,w$) holds since we can use the first valid block to
    allocate $f$ on $w$. The complexity of checking whether a block is valid
    corresponds
  to a linear-time check of the list of invalidation policies associated with
  that block.
  Since there is a linear number of blocks w.r.t. the size of the input, then
  the reachability problem is linear in the size of the input.
  \qed
\end{proof}
We use the same proof technique and show that co-occurrence has linear time
complexity.

\begin{theorem}
  \label{theorem:co-occurence}
  \CoOccur($p, \reg, C,\{f,g\},w$) has linear time complexity.\end{theorem}
\vspace{-4mm}

\begin{proof}
  We adopt the same proof structure as that of \cref{theorem:poly_reach}.

  We follow the same proof technique of \cref{lemma:only_target} and 
  derive that, assuming an empty configuration $C$,
  \CoOccur($p,\reg,C,\{f,g\},w$) holds if and only if
  \CoOccur($\simple(p,w),C,\{f,g\},w$) holds.

  Following the proof of \cref{theorem:poly_reach}, we reduce
  \CoOccur($p,\reg,C,\{f,g\},w$) to the same co-occurrence problem where the
  policy is $\simple(p,w)$ and $C$ is empty.
  We can solve this simplified problem by verifying whether there exists a
  sequence of allocations such that we can schedule $f$ and $g$ on $w$.
  We verify this by checking twice each block relative to $f$ and $g$, i.e., we
  check whether the block is valid either assuming no function scheduled on $w$
  or that one instance of the other function has been scheduled. If we
  have that one of the two functions has at least one block which is initially
  valid, and the other function has at least one block which is still valid after
  the allocation of the other function on \(w\), then we
  can conclude that \CoOccur($\simple(p,w),\reg,C,\{f,g\},w$) holds and, thereby, also \CoOccur($p,\reg,C,\{f,g\},w$) holds.
  Indeed, the potential allocation of other functions cannot allow for the
  co-occurrence of $f$ and $g$ on $w$ for the anti-monotonicity of the {\app} capacity constraints.

  Since the verification procedure described above performs two times a linear-time check, the complexity of checking co-occurence is linear.
  \qed
\end{proof}

\subsection{On the Support for Affinity-awareness of \app}
\label{subsec:support}
The linear time complexity of co-occurrence is a positive result, e.g., we can
efficiently check possible violations of security properties of scripts, like
the co-location of ``trusted'' and ``untrusted'' functions.

However, this simplicity comes at a cost.
We can avoid the co-occurrence of functions only by using strong
resource-sharing limitations.
In the proof of \cref{theorem:co-occurence}, we have observed that we can check
\CoOccur($p,\reg,C,\{f,g\},w$) by verifying the following property on the
simplified policy $\simple(p,w)$:
\emph{``one of the two functions has at least one block which is initially
  valid, and the other function has at least one block which is still valid after
  the allocation of the other function on \(w\)''}.
There are two possible ways to make such property false.
One possibility is to avoid that \emph{``one of the two functions has at least
  one block which is initially valid''}, i.e., one between
\Reach($\simple(p,w),C,f,w$)
and \Reach($\simple(p,w),C,g,w$) must be false. By \cref{lemma:only_target}, making either predicate false means that the \app script eliminates
$w$ from the workers ever available for the scheduling of $f$ or $g$.
If $w$ is among the workers available for the scheduling of both $f$ and $g$,
the other possibility to make the above property false
is to avoid that one of the two functions
\emph{``has at least one block which is still valid after
  the allocation of the other function on $w$''}.
We can make all blocks invalid for one of the two functions by artificially limiting the capacity or the number of functions we can
allocate on a worker (i.e., setting the \hlopt{capacity\_used} below the
cumulative occupancy of the two functions or setting
\hlopt{max\_concurrent\_invocations} to 1).
These limitations lead us to deem \app unfit to effectively express
anti-affinity constraints.

The situation for affinity, i.e., that the
scheduler allocates a given function only on a worker that hosts its affine functions,
is even poorer.
Consider the
transitions $C \xRightarrow{(start,f,w)} C'$, corresponding to the
scheduling of function $f$ on a worker $w$. If $f$ is expected to be
affine with another function $g$, we consider this transition possible
only if the worker $w$ already hosts the function $g$ in the configuration $C$.
Consider now the configuration $C''$, obtained by removing all occurrences
of $g$ from the worker $w$. We have that
the schedule of $f$
can be trivially performed also in $C''$ since 
$w$ has less load (thus, less \hlopt{capacity\_used}) and contains fewer functions (thus, a smaller number for \hlopt{max\_concurrent\_invocations}).
Hence, $f$ can be scheduled on $w$ even if $g$ is absent and therefore \app cannot enforce affinity constraints.

\section{Reachability and Co-Occurence in \appp}
\label{sec:appp}

The formal analysis of the properties discussed in \cref{sec:app_properties}
shows that \app is not expressive enough to capture anti-affinity and affinity
constraints
and that the reachability and
co-occurrence problems have linear time complexity. In this section 
 we prove that for {\appp} such problems become instead PSPACE-complete.
On the one hand, from the standpoint of reachability verification, we can see
this strong complexity increase as problematic. On the other hand, this result
formalises the increment in language expressiveness, i.e., that {\appp} can
specify more sophisticated scheduling policies than \app.

\begin{theorem}\label{theo:PSPACE}
In {\appp} the problem \Reach($p,reg,C,f,w$) is PSPACE-complete.
\end{theorem}
\vspace{-4mm}

\begin{proof}
The problem is in PSPACE since it is possible to store the configuration in a
matrix with a row for each worker, a column for each function, and the number of
instances of a function on a worker stored in the corresponding cell. The
occupation in memory of such matrix is polynomial in the size of the input,
given that the input contains the description of all the functions (in $reg$) as
well as the capacity of all the workers (included in the initial configuration
$C$). Hence, we can implement a nondeterministic algorithm that checks
reachability using only polynomial space. The result follows from the
coincidence between PSPACE and NPSPACE \cite{Savitch}.

We prove the problem to be PSPACE-hard by reduction from the PLANSAT problem, which is the problem of determining the existence of a solution for
propositional planning where:
\begin{itemize} 
\item we represent the initial state by a finite set of ground atomic conditions
considered 
true;
\item we can use operators to change the current state: an operator first checks
some positive and negative pre-conditions (the positive ones should hold, the
negative ones should not hold) and then modifies the state depending on some
positive and negative post-conditions (the positive conditions become true, the
negative become false);
\item we represent the goals by the conjunction of some positive and negative
conditions; the positive conditions should hold in the final state, while the
negative ones should not hold.
\end{itemize}
For simplicity, we impose operators to have only one post-condition. This
restricted version of PLANSAT was proved to be PSPACE-hard by Bylander
\cite{DBLP:journals/ai/Bylander94}.
%
%
%
%
%
The idea behind the encoding of PLANSAT is to consider, for each condition $c$,
the functions $c+$ and $c-$ that are respectively scheduled to indicate whether
the condition $c$ is currently true (i.e. positive) or false (i.e. negative).
For all the operators $o$, we consider a function $f_o$ that can be scheduled
only if the pre-conditions hold. Only while a function $f_o$ is scheduled, the
function corresponding to its post-condition $c+$ (resp. $c-$) can be scheduled
provided that no function corresponding to its negation $c-$ (resp. $c+$) are
also scheduled. We assume the availability of only one worker $w$, with a
capacity sufficient to schedule at least one instance of all the functions $c+$,
$c-$, $f_o$, plus the auxiliary functions $start$ and $goal$ described below.

The auxiliary function $start$ is anti-affine to all the other functions. For
the condition functions $c+$ and $c-$ we consider several possible blocks in
their scheduling policy. Concerning the encoding of the initial state we proceed
as follows: for each condition $c$ which holds in the initial state, we assume a
scheduling block for the function $c+$ that imposes affinity with $start$, while
for all the other conditions $d$ which do not hold in the initial state, we
assume a block for $d-$ that imposes affinity with $start$. Naturally, $c+$ is
antiaffine to $c-$ and vice versa.

The function $f_o$ is affine 
with the $c+$ functions corresponding to the positive pre-conditions of the
operation $o$, with the $c-$ functions corresponding to the negative
pre-conditions of $o$, and is anti-affine with $start$ and all the operator
functions $f_{p}$ (for every operator $p$).
Moreover, if the operator $o$ has a positive (resp. negative) post-condition
$c$, we add a block to the scheduling policy of the corresponding condition
function $c+$ (resp. $c-$) that imposes affinity with $f_{o}$ and anti-affinity
with $c-$ (resp. $c+$) and with the $start$ function.

Finally, the auxiliary function $goal$ is affine with all the functions $c+$,
for all the positive conditions $c$ in the goals, and the functions $d-$, for
all the negative conditions $d$, and it is anti-affine with $start$.

We have that $goal$ can be scheduled on the worker $w$ (assuming $w$ empty in
the initial configuration) if and only if there exists a solution to the
considered propositional planning problem.

The right-to-left implication follows from the existence of a plan, i.e., a
sequence of operators that can change the initial state in a final state
satisfying the goals. We can consider a sequence of function scheduling that
initially schedules $start$, all the positive pre-condition functions $c+$ and
the negative pre-condition functions $c-$ of the initial state, and then removes
$start$. Then the functions $f_o$ can be scheduled following the order of
execution of the corresponding operations in the plan. 
While a function $f_o$ is scheduled, also the function corresponding to its
post-condition is scheduled. In the reached configuration, the $goals$ function
can be finally scheduled.

The left-to-right implication follows from the following observations. The
unique function that can be initially scheduled in an empty worker is $start$
because all the other functions impose affinity constraints. The unique
functions that can be scheduled while $start$ is on $w$, are the $c+$ and $c-$
functions representing the initial state. When $start$ is removed we have that:
if a function $c+$ is on $w$, then $c$ is a condition in the initial state,
while if $c-$ is on $w$, then $c$ is not in the initial state. This formalises
our notion of correctness of the encoding of the state: the conditions functions
currently present on $w$ represent a correct (possibly partial) state of the
planning system in that if $c+$ is on $w$, the condition $c$ holds in that
state, while if $c-$ is on $w$, the condition $c$ does not hold. Such a form of
correctness is preserved during the possible function scheduling sequences. In
fact, only functions $f_o$ corresponding to operations $o$ whose pre-conditions
are satisfied can be scheduled, and only the functions $c+$ or $c-$
corresponding to the post-condition of such operations can be scheduled
(assuming that the function corresponding to its negation has been removed from
$w$). The other modifications that can occur coincide with the removal of
functions from $w$, but these do not break our correctness condition. Hence, if
the function $goal$ can be scheduled on $w$, this implies that it is possible to
reach, in the planning system, a state in which the positive conditions imposed
by the goals hold, while the negative conditions do not hold.
\qed
\end{proof}

\begin{corollary}\label{cor:co_occur_PSPACE} In {\appp} the problem \CoOccur($p,
reg, C,\{f,g\},w$) is PSPACE-complete.
\end{corollary}
\vspace{-4mm}

\begin{proof}
We use the same argument as in \Cref{theo:PSPACE} (i.e., we can store a
configuration in polynomial space) to prove that also the co-occurrence problem
is in PSPACE. Concerning PSPACE-completeness, we can observe that the
co-occurrence problem cannot be computationally less complex than reachability.
In fact, given an instance of the reachability problem for a function $f$ and a
worker $w$, we can reduce such a problem to the co-occurrence on $w$ of the
function $f$ and an auxiliary function $g$ that we can schedule only on $w$,
even if $w$ contains already one copy of $f$.
\qed
\end{proof}

\section{Reachability and Co-Occurence with Polarised \appp Affinities}
\label{sec:appp_other}
In this section we
investigate \emph{polarised} fragments of \appp, where one can express either
affinity or anti-affinity but not mix these within the same script. 
Practically, this study aims to understand if we can use algorithmically
tractable analyses with fragments of \appp, e.g., when a script uses only the
syntax of a polarised fragment.

%
More precisely, we consider a \emph{positively-polarised} fragment of \appp that
contains only affinity constraints, and a \emph{negatively-polarised} fragment
where users can only express anti-affinity.
We study the complexity of the reachability and co-occurrence problems for the
two fragments. We show that 
in positively-polarised \appp
the problems are NP-hard (hence not tractable unless P=NP), while in
negatively-polarised \appp both
problems have linear time complexity.
%

First, we prove that in positively-polarised {\appp}---where
scripts carry only affinity constraints of the form $\textit{id}$
---the reachability problem
is still NP-hard.

\begin{theorem}\label{theo:PSPACE_pp} In positively-polarised {\appp}, i.e.,
where one can only express affinity constraints, the problems
\Reach($p,reg,C,f,w$) and \CoOccur($p, reg, C,\{f,g\},w$) are NP-hard.
\end{theorem}

\noindent
\emph{Proof sketch} (full proof available in \cref{appendix-theorems}).

The NP-hard result is proved by reduction from 3SAT~\cite{DBLP:conf/stoc/Cook71}, i.e., checking the satisfiability of a
boolean formula where each clause has at most three
litarals. The idea is to encode each possible literal with a
function $l_i$ that can be scheduled and use the capacity
constraints to limit the possibility of scheduling both $l_i$ and its negation
$!l_i$. Each clause is instead encoded with a corresponding function $c_j$ that can be scheduled only if at least one of its literal functions $l_j^1,l_j^2,l_j^3$ has been already scheduled.

To limit the possibility that only one function can be scheduled at a time, we consider a unique worker $w$ with capacity $3*2^{n+2}-2$, where $n$ is the number of variables in the 3SAT formula. For each variable $x_i$ we consider five functions $x_i$, $!x_i$, $left_i$,
$mid_i$, $right_i$ all consuming $2^i$ resources, and with $x_i$, $!x_i$
having max capacity $(4 * 2^i-1) * 100 / (3*2^{n+2}-2)$. Moreover, we impose
$x_i$ affine to $left_i$ and $mid_i$, and $!x_i$ affine to $mid_i$ and
$right_i$. In this way, it is not possible to have the functions $x_i$
and $!x_i$ both scheduled on the worker $w$ because this requires to
schedule at the same time $x_i$,  $!x_i$, $mid_i$ and one between $left_i$
and $right_i$, violating the capacity constraint.

For each clause, we consider one function $c_j$ consuming $2^{n+2}$ resources.
Each function $c_j$ has three scheduling blocks. For the first function $c_1$
the first block imposes $l_1^1$ as an affine function, the second one considers
$l_1^2$, while the third one $l_1^3$. For $j>1$, each scheduling block imposes
an affinity with also the function $c_{j-1}$, besides the corresponding
$l_j^1,l_j^2,l_j^3$ literal functions, respectively.
\qed

\vspace{5pt}
Then, we prove that, in the negatively-polarised fragment of \appp---where
scripts carry only anti-affinity constraints of the form $\textit{!id}$
---both the reachability and
co-occurrence problems are linear w.r.t. the size of the input.

\begin{theorem}\label{theo:anti-affinity} In negatively-polarised {\appp}, i.e.,
{\appp} with 
only 
anti-affinity constraints, 
the problems
\Reach($p,reg,C,f,w$) and \CoOccur($p, reg, C,\{f,g\},w$) have linear time complexity.
\end{theorem}
\vspace{-5mm}

\begin{proof}
We first observe that anti-affinity constraints are 
anti-monotonic, i.e., if a worker satisfies some anti-affinity
constraints, the same worker with a smaller load (i.e. less functions)
satisfies those constraints too.
In the light of this property of anti-affinity
all the steps of the proofs of \Cref{lemma:only_target},
\Cref{theorem:poly_reach}, and \Cref{theorem:co-occurence} hold also for
the negatively-polarised fragment of {\appp}. Thus, the reachability and
co-occurrence problems remain linear in negatively-polarised \appp.
\qed
%
%
%
\end{proof}

\section{Related Work and Discussion}

To the best of our knowledge, this is the first presentation of a formal model
to reason on the semantics of serverless function scheduling.

Broadening our scope, the works we see the closest to ours come from the
neighbouring area of microservices---the state-of-the-art style for cloud
architectures~\cite{DGLMMMS17}. Proposals in this direction are by Baarzi and
Kesidis~\cite{BK21}, who present a framework for the deployment of microservices
that infers and assigns affinity and anti-affinity traits to microservices to
orient the distribution of resources and microservices replicas on the available
machines; Sampaio et al.~\cite{SRBR19}, who introduce an adaptation mechanism
for microservice deployment based on microservice affinities (e.g., the more
messages microservices exchange the more affine they are) and resource usage;
Sheoran et al.~\cite{SFSM21}, who propose an approach that computes procedural
affinity of communication among microservices to make placement decisions.

Looking at the industry, Azure Service Fabric~\cite{web:azure_sf} provides a
notion of \emph{service affinity} that ensures the placement of replicas of a
service on the same nodes as those of another, affine service. Another example
is Kubernetes, which has a notion of \emph{node affinity} and \emph{inter-pod
(anti-)affinity} to express advanced scheduling logic for the optimal
distribution of pods~\cite{web:kube_affinity}. Overall, the mentioned work
proves the usefulness of affinity-aware deployments at lower layers than FaaS
(e.g., VMs, containers, microservices).

Regarding the constructs we have proposed for expressing (anti-)affinity
constraints in \appp, we observe that an alternative approach could be to let
the user directly declare the properties to enforce, leaving the platform to
realise them at run time. For instance, we could have added support for
quantified formulas delegating to 
the scheduling runtime of this \app variant to allocate a function only if the
allocation satisfies the formula or fail otherwise. The problem with this
approach is scalability.
Indeed, checking the satisfiability of a property's formula may take an
exponential time on the size of the formula, workers, and functions. Contrarily,
the \appp scheduler allocates a function on a worker (if any) in liner time on
the size of the \appp script and workers.

%

A recent trend of FaaS is the definition/handling of the composition/workflows
of functions, like AWS step-functions~\cite{web:step_functions} and Azure
Durable functions~\cite{BGJKMM21}. The main idea behind these works is to allow
users to define workflows as the composition of functions with their branching
logic, parallel execution, and error handling. The orchestrator/controller of
the platform then uses the workflow to manage function executions and handle
retries, timeouts, and errors. Our proposal is orthogonal to these works.
Indeed, assuming the workflow is available, the orchestrator developed for
handling serverless workflows should be extensible with an \appp-like script to
specify where to schedule the functions within a given workflow. Future work on
this integration would support the enforcement of even more expressive policies
than \appp, like preventing function instances of the same workflow from sharing
nodes.

In the literature on serverless, we can spot proposals that might adopt affinity
and anti-affinity relations to encode and complement desirable properties of
function scheduling, otherwise implemented via ad-hoc, platform-wide policies.
For instance, some works present serverless architectures that enable the
efficient composition of functions co-located on the same
host~\cite{ACRSSBAH18,SP20,SRBFC21}. Here, \appp can help to parametrise the
co-location of functions, expressed in terms of affinity and anti-affinity
constraints. Another example regards security, e.g., Pubali et
al.~\cite{PPTMAA20} present a serverless platform where developers can
constrain the information flow among functions to avoid attacks due to container
reuse and data exfiltration. In this case, \appp can complement flow policies
with affinity and anti-affinity constraints that restrict the co-tenancy of
functions and their flows of communication.
Another interesting proposal, Palette~\cite{AGLFCGBBF23}, uses optional opaque
parameters in function invocations to inform the load balancer of Azure
Functions on the affinity with previous invocations and the data they
produced. While Palette does not support (anti-)affinity constraints, it allows
users to express which invocations benefit from running on the same node. We
deem an interesting future work extending \appp to support a notion of
(anti-)affinity that considers the history of scheduled functions.

\section{Conclusion and Future Work}
\label{sec:conclusion}
%
%
We started covering the ground for a rigorous treatment of FaaS
scheduling, focusing on affinity.
We have first performed a formal analysis of \app, a platform-agnostic language 
for FaaS scheduling policies, showing that it is unfit for expressing
interesting (anti-)affinity constraints between functions.
These initial results motivated the definition of an extension of the
\app language, which we call \appp, with constructs for expressing 
(anti-)affinity constraints. We formally prove the increment of expressiveness
of the language by showing that affinity-aware reachability problems, 
having linear time complexity in \app, turn out to be PSPACE-complete
in \appp. The proof of PSPACE-hardness is by reduction from PLANSAT,
a well-known planning problem.

The table below summarises the support for affinity policies and the complexity bounds of the co-occurrence (and reachability) problem of \app, \appp, and the latter's affinity-only and anti-affinity-only fragments.

\begin{center}
\begin{adjustbox}{width=.8\textwidth} 
\begin{tabular}{|c|c|c|c|c|}
\hline
& \app & neg. polarised \appp & pos. polarised \appp & \appp\\
\hline
\hline
Affinity & \xmark & \xmark & \cmark & \cmark \\
\hline
Anti-affinity & \xmark & \cmark & \xmark & \cmark \\
\hline
Lowerbound & Linear & Linear & NP & PSpace \\
\hline
Upperbound & Linear & Linear & PSpace & PSpace \\
\hline
\end{tabular}
\end{adjustbox}
\end{center}


Directions for future work include capturing
configurations whose workers can change, i.e., when workers (dis)appear while
the platform is running. In such dynamic scenarios, the configuration space is
infinite, which could have repercussions on the complexity/decidability
of the reachability and co-occurrence problems.
%
We also plan to define a timed 
semantics for \appp{}, e.g., to specify the expected execution time of functions
on workers. One such model could support quantitative analysis, e.g., to
estimate the completion time of a FaaS application or the distribution of
workers' load over time, i.e., it would allow us to quantitatively reason on
policies, e.g., whether they could lead to bad performance or underutilisation
of workers.

\bibliography{biblio}

\begin{thebibliography}{10}
\providecommand{\url}[1]{\texttt{#1}}
\providecommand{\urlprefix}{URL }
\providecommand{\doi}[1]{https://doi.org/#1}

\bibitem{ABE18}
Abad, C.L., Boza, E.F., Eyk, E.V.: Package-aware scheduling of faas functions.
  In: Proc. of {ACM/SPEC} {ICPE}. pp. 101--106. {ACM} (2018).
  \doi{10.1145/3185768.3186294}

\bibitem{AGLFCGBBF23}
Abdi, M., Ginzburg, S., Lin, X.C., Faleiro, J., Chaudhry, G.I., Goiri, I.,
  Bianchini, R., Berger, D.S., Fonseca, R.: Palette load balancing: Locality
  hints for serverless functions. In: Proceedings of the Eighteenth European
  Conference on Computer Systems. pp. 365--380 (2023)

\bibitem{ACRSSBAH18}
Akkus, I.E., Chen, R., Rimac, I., Stein, M., Satzke, K., Beck, A., Aditya, P.,
  Hilt, V.: $\{$SAND$\}$: Towards $\{$High-Performance$\}$ serverless
  computing. In: 2018 Usenix Annual Technical Conference (USENIX ATC 18). pp.
  923--935 (2018)

\bibitem{AFFRSSW18}
Alpernas, K., Flanagan, C., Fouladi, S., Ryzhyk, L., Sagiv, M., Schmitz, T.,
  Winstein, K.: Secure serverless computing using dynamic information flow
  control. Proceedings of the ACM on Programming Languages  \textbf{2}(OOPSLA),
   1--26 (2018)

\bibitem{web:azurefunctions}
Azure, M.: Microsoft azure functions. \url{https://azure.microsoft.com/} (11
  2022)

\bibitem{web:azure_sf}
Azure, M.: Microsoft azure functions.
  \url{https://learn.microsoft.com/en-us/azure/service-fabric/service-fabric-overview}
  (11 2022)

\bibitem{BK21}
Baarzi, A.F., Kesidis, G.: Showar: Right-sizing and efficient scheduling of
  microservices. In: Proceedings of the ACM Symposium on Cloud Computing. pp.
  427--441 (2021)

\bibitem{BCCCFIMMRS17}
Baldini, I., Castro, P., Chang, K., Cheng, P., Fink, S., Ishakian, V.,
  Mitchell, N., Muthusamy, V., Rabbah, R., Slominski, A., et~al.: Serverless
  computing: Current trends and open problems. In: Research advances in cloud
  computing, pp. 1--20. Springer (2017)

\bibitem{BS21}
Banaei, A., Sharifi, M.: Etas: predictive scheduling of functions on worker
  nodes of apache openwhisk platform. The Journal of Supercomputing  (9 2021).
  \doi{10.1007/s11227-021-04057-z}

\bibitem{BQ21}
Baresi, L., Quattrocchi, G.: Paps: A serverless platform for edge computing
  infrastructures. Frontiers in Sustainable Cities  \textbf{3},  690660 (2021)

\bibitem{BGJKMM21}
Burckhardt, S., Gillum, C., Justo, D., Kallas, K., McMahon, C., Meiklejohn,
  C.S.: Durable functions: semantics for stateful serverless. Proceedings of
  the ACM on Programming Languages  \textbf{5}(OOPSLA),  1--27 (2021)

\bibitem{DBLP:journals/ai/Bylander94}
Bylander, T.: The computational complexity of propositional {STRIPS} planning.
  Artif. Intell.  \textbf{69}(1-2),  165--204 (1994).
  \doi{10.1016/0004-3702(94)90081-7},
  \url{https://doi.org/10.1016/0004-3702(94)90081-7}

\bibitem{CAVJLLPPR20}
Casale, G., Arta{\v{c}}, M., Van Den~Heuvel, W.J., van Hoorn, A., Jakovits, P.,
  Leymann, F., Long, M., Papanikolaou, V., Presenza, D., Russo, A., et~al.:
  Radon: rational decomposition and orchestration for serverless computing.
  SICS Software-Intensive Cyber-Physical Systems  \textbf{35}(1),  77--87
  (2020)

\bibitem{web:googlefunctions}
Cloud, G.: Google cloud functions. \url{https://cloud.google.com/functions/}
  (11 2022)

\bibitem{DBLP:conf/stoc/Cook71}
Cook, S.A.: The complexity of theorem-proving procedures. In: Harrison, M.A.,
  Banerji, R.B., Ullman, J.D. (eds.) Proceedings of the 3rd Annual {ACM}
  Symposium on Theory of Computing, May 3-5, 1971, Shaker Heights, Ohio, {USA}.
  pp. 151--158. {ACM} (1971). \doi{10.1145/800157.805047},
  \url{https://doi.org/10.1145/800157.805047}

\bibitem{PPTMAA20}
Datta, P., Kumar, P., Morris, T., Grace, M., Rahmati, A., Bates, A.: Valve:
  Securing function workflows on serverless computing platforms. In:
  Proceedings of The Web Conference 2020. pp. 939--950 (2020)

\bibitem{DGMTZ22}
{De Palma}, G., Giallorenzo, S., Mauro, J., Trentin, M., Zavattaro, G.: A
  declarative approach to topology-aware serverless function-execution
  scheduling. In: {IEEE} International Conference on Web Services, {ICWS} 2022,
  Barcelona, Spain, July 10-16, 2022. pp. 337--342. {IEEE} (2022).
  \doi{10.1109/ICWS55610.2022.00056}

\bibitem{PGMZ20}
{De Palma}, G., Giallorenzo, S., Mauro, J., Zavattaro, G.: Allocation priority
  policies for serverless function-execution scheduling optimisation. In:
  Service-Oriented Computing - 18th International Conference, {ICSOC} 2020,
  Dubai, United Arab Emirates, December 14-17, 2020, Proceedings. Lecture Notes
  in Computer Science, vol. 12571, pp. 416--430. Springer (2020).
  \doi{10.1007/978-3-030-65310-1\_29}

\bibitem{DGLMMMS17}
Dragoni, N., Giallorenzo, S., Lluch{-}Lafuente, A., Mazzara, M., Montesi, F.,
  Mustafin, R., Safina, L.: Microservices: Yesterday, today, and tomorrow. In:
  Present and Ulterior Software Engineering, pp. 195--216. Springer (2017).
  \doi{10.1007/978-3-319-67425-4\_12}

\bibitem{web:fission}
Fission: Fission. \url{https://fission.io/} (11 2022)

\bibitem{HSHVAA16}
Hendrickson, S., Sturdevant, S., Harter, T., Venkataramani, V., Arpaci-Dusseau,
  A.C., Arpaci-Dusseau, R.H.: Serverless computation with openlambda. In: 8th
  $\{$USENIX$\}$ Workshop on Hot Topics in Cloud Computing (HotCloud 16) (2016)

\bibitem{ZE21}
Jia, Z., Witchel, E.: Boki: Stateful serverless computing with shared logs. In:
  Proc. of ACM SIGOPS SOSP. pp. 691--707. ACM, New York, NY, USA (2021).
  \doi{10.1145/3477132.3483541}

\bibitem{Jonas-etal:BerkeleyViewOnServerless}
Jonas, E., Schleier-Smith, J., Sreekanti, V., Tsai, C.C., Khandelwal, A., Pu,
  Q., Shankar, V., Menezes~Carreira, J., Krauth, K., Yadwadkar, N., Gonzalez,
  J., Popa, R.A., Stoica, I., Patterson, D.A.: Cloud programming simplified: A
  berkeley view on serverless computing. Tech. Rep. UCB/EECS-2019-3, EECS
  Department, University of California, Berkeley (02 2019)

\bibitem{KSB19}
Kehrer, S., Scheffold, J., Blochinger, W.: Serverless skeletons for elastic
  parallel processing. In: 2019 IEEE 5th International Conference on Big Data
  Intelligence and Computing (DATACOM). IEEE. pp. 185--192 (2019)

\bibitem{KGB20}
Kelly, D., Glavin, F., Barrett, E.: Serverless computing: Behind the scenes of
  major platforms. In: 2020 IEEE 13th International Conference on Cloud
  Computing (CLOUD). pp. 304--312. IEEE (2020)

\bibitem{SAVA21}
Kotni, S., Nayak, A., Ganapathy, V., Basu, A.: Faastlane: Accelerating
  function-as-a-service workflows. In: Proc. of {USENIX} {ATC}. pp. 805--820.
  {USENIX} Association (2021)

\bibitem{web:kube_affinity}
Kubernetes: Assign pods to nodes using node affinity.
  \url{https://kubernetes.io/docs/tasks/configure-pod-container/assign-pods-nodes-using-node-affinity/}
  (11 2022)

\bibitem{KNJ18}
Kuntsevich, A., Nasirifard, P., Jacobsen, H.A.: A distributed analysis and
  benchmarking framework for apache openwhisk serverless platform. In: Proc. of
  Middleware (Posters). pp.~3--4 (2018)

\bibitem{service_fabric_affinity}
Microsoft: Configuring and using service affinity in service fabric.
  \url{https://learn.microsoft.com/en-us/azure/service-fabric/service-fabric-cluster-resource-manager-advanced-placement-rules-affinity}
  (2023)

\bibitem{AHKSNV19}
Mohan, A., Sane, H., Doshi, K., Edupuganti, S., Nayak, N., Sukhomlinov, V.:
  Agile cold starts for scalable serverless. In: Proc. of HotCloud 19. {USENIX}
  Association, Renton, WA (jul 2019)

\bibitem{OYZHHAA18}
Oakes, E., Yang, L., Zhou, D., Houck, K., Harter, T., Arpaci-Dusseau, A.,
  Arpaci-Dusseau, R.: $\{$SOCK$\}$: Rapid task provisioning with
  $\{$Serverless-Optimized$\}$ containers. In: 2018 USENIX Annual Technical
  Conference (USENIX ATC 18). pp. 57--70 (2018)

\bibitem{web:openfaas}
OpenFaaS: Openfaas. \url{https://www.openfaas.com/} (11 2022)

\bibitem{web:openstack_affinity}
OpenStack: Openstack documentation, affinity.
  \url{https://docs.openstack.org/project-deploy-guide/openstack-ansible/ocata/app-advanced-config-affinity.html}
  (03 2019)

\bibitem{web:OpenWhisk}
OpenWhisk, A.: Apache openwhisk. \url{https://openwhisk.apache.org/} (11 2022)

\bibitem{PalmaGMTZ23}
Palma, G.D., Giallorenzo, S., Mauro, J., Trentin, M., Zavattaro, G.: Formally
  verifying function scheduling properties in serverless applications. {IT}
  Prof.  \textbf{25}(6),  94--99 (2023). \doi{10.1109/MITP.2023.3333071},
  \url{https://doi.org/10.1109/MITP.2023.3333071}

\bibitem{SRBFC21}
Sabbioni, A., Rosa, L., Bujari, A., Foschini, L., Corradi, A.: A shared memory
  approach for function chaining in serverless platforms. In: 2021 IEEE
  Symposium on Computers and Communications (ISCC). pp.~1--6. IEEE (2021)

\bibitem{SRBR19}
Sampaio, A.R., Rubin, J., Beschastnikh, I., Rosa, N.S.: Improving
  microservice-based applications with runtime placement adaptation. Journal of
  Internet Services and Applications  \textbf{10}(1),  1--30 (2019)

\bibitem{SSGP17}
Samp\'{e}, J., S\'{a}nchez-Artigas, M., Garc\'{\i}a-L\'{o}pez, P., Par\'{\i}s,
  G.: Data-driven serverless functions for object storage. In: Middleware. pp.
  121--133. Middleware ’17, ACM (2017). \doi{10.1145/3135974.3135980},
  \url{https://doi.org/10.1145/3135974.3135980}

\bibitem{Savitch}
Savitch, W.J.: Relationships between nondeterministic and deterministic tape
  complexities. J. Comput. Syst. Sci.  \textbf{4}(2),  177--192 (1970).
  \doi{10.1016/S0022-0000(70)80006-X},
  \url{https://doi.org/10.1016/S0022-0000(70)80006-X}

\bibitem{web:IntroducingAwsLambda}
Services, A.W.: Introducing aws lambda.
  \url{https://aws.amazon.com/about-aws/whats-new/2014/11/13/introducing-aws-lambda/}
  (11 2022)

\bibitem{web:step_functions}
Services, A.W.: Aws step functions.
  \url{https://aws.amazon.com/step-functions/} (07 2023)

\bibitem{SBW19}
Shahrad, M., Balkind, J., Wentzlaff, D.: Architectural implications of
  function-as-a-service computing. In: Proc. of {MICRO}. pp. 1063--1075 (2019)

\bibitem{SFGCBCLTRB20}
Shahrad, M., Fonseca, R., Goiri, {\'I}., Chaudhry, G., Batum, P., Cooke, J.,
  Laureano, E., Tresness, C., Russinovich, M., Bianchini, R.: Serverless in the
  wild: Characterizing and optimizing the serverless workload at a large cloud
  provider. In: Proc. of {USENIX} {ATC}. pp. 205--218 (2020)

\bibitem{SFSM21}
Sheoran, A., Fahmy, S., Sharma, P., Modi, N.: Invenio: Communication affinity
  computation for low-latency microservices. In: Proceedings of the Symposium
  on Architectures for Networking and Communications Systems. pp. 88--101
  (2021)

\bibitem{SP20}
Shillaker, S., Pietzuch, P.: Faasm: Lightweight isolation for efficient
  stateful serverless computing. In: Proc. of {USENIX} {ATC}. pp. 419--433.
  {USENIX} Association (2020)

\bibitem{SFP20}
Silva, P., Fireman, D., Pereira, T.E.: Prebaking functions to warm the
  serverless cold start. In: Proc. of Middleware. pp. 1--13. Middleware '20,
  ACM, New York, NY, USA (2020). \doi{10.1145/3423211.3425682}

\bibitem{SJCGB22}
Smith, C.P., Jindal, A., Chadha, M., Gerndt, M., Benedict, S.: Fado: Faas
  functions and data orchestrator for multiple serverless edge-cloud clusters.
  In: 2022 IEEE 6th International Conference on Fog and Edge Computing (ICFEC).
  pp. 17--25. IEEE (2022)

\bibitem{SA20}
Solaiman, K., Adnan, M.A.: Wlec: A not so cold architecture to mitigate cold
  start problem in serverless computing. In: 2020 IEEE International Conference
  on Cloud Engineering (IC2E). pp. 144--153 (2020).
  \doi{10.1109/IC2E48712.2020.00022}

\bibitem{SWLSGHT20}
Sreekanti, V., Wu, C., Lin, X.C., Schleier-Smith, J., Gonzalez, J.E.,
  Hellerstein, J.M., Tumanov, A.: Cloudburst: Stateful functions-as-a-service.
  Proc. VLDB Endow.  \textbf{13}(12),  2438--2452 (Jul 2020).
  \doi{10.14778/3407790.3407836}

\bibitem{SG19}
Suresh, A., Gandhi, A.: Fnsched: An efficient scheduler for serverless
  functions. In: {WOSC@Middleware}. pp. 19--24. {ACM} (2019).
  \doi{10.1145/3366623.3368136}

\bibitem{WLZRS18}
Wang, L., Li, M., Zhang, Y., Ristenpart, T., Swift, M.: Peeking behind the
  curtains of serverless platforms. In: 2018 USENIX Annual Technical Conference
  (USENIX ATC 18). pp. 133--146 (2018)

\bibitem{web:YAML}
YAML: {YAML} specification. \url{https://yaml.org/spec/} (11 2022)

\end{thebibliography}

\newpage
\appendix

\section{Semantics}
\label{appendix-semantics}

We start by presenting the semantics of \app and then extend it to define the
behaviour of the (anti-)affinity constraints, obtaining the semantics of \appp.

\subsection{\app semantics}

While \app scripts are
YAML-compliant, we slightly stylise the \app syntax to increase readability (for
instance, we omit quotes around strings, e.g., \hlopt{*} instead of
\hlopt{"*"}). For simplicity, we also assume two minor variations w.r.t.~\cite{PGMZ20}:
i) we avoid modelling
\hln{strategy}\hlstr{:} \hlopt{platform} and \hln{invalidate}\hlstr{:}
\hlopt{overload} since they are platform-specific
(e.g., in Apache OpenWhisk \hlopt{platform} implies the usage of its hardcoded
strategy, based on a co-prime heuristic selection logic~\cite{PGMZ20}); ii)
we rename the \hlopt{random} strategy option to \hlopt{any}, to point out that
it performs non-deterministic worker selection instead of a
uniformly-distributed one, as \hlopt{random} does.
We assume that \app scripts \emph{a}) include the policy
``$\mathtt{default}\hlstr{:}\ \texttt{-}\ \hln{workers}\ :\ \hlopt{*}$'' unless
the user customises the \texttt{default} policy and \emph{b}) show the \texttt{default} policy as the last one in order of appearance.
Users can customise the default policy, except it can only
\hlopt{fail}, to avoid \hln{followup} loops. Any policy that leaves
\hln{strategy} or \hln{invalidate} unspecified allocates functions on
\hlopt{any} worker to 100\% \hlopt{capacity\_used}.

We define the behaviour of \app scripts as a labelled transition system (LTS)
operational semantics. In the definition of the LTS, we use these domains, structures, and
functions:
\vspace{-1em}

\vspace{-1em}\begin{align*}
  w \in \Works \subset \mathit{Identifiers}
  & &
  f \in \Funcs
  & & 
  C \in \mathcal C \triangleq \Works \to \mathit{Multiset}(\Funcs) \times \mathbb{N} \times \mathbb{N}
  \\
  t \in \Tags \subset \mathit{Identifiers}
  &  &
  \reg \in \Funcs \to \mathbb{N} \times \Tags
  & &
  \enc{\cdot}\ :\ app \to \Policies
  \\
  \policy \in \Policies \triangleq \Tags \to \mathit{List}(\Blocks)
  & &
  & &
  \block \in \Blocks \triangleq (\mathit{List}(\Works) \cup \hlopt{*}) \times \textit{\sopt} \times \mathit{List}(\textit{\iopt})
\end{align*}

We use $\Works$, ranged over by $w$, to denote the set of workers, while
$\Funcs$, ranged over by $f$, denotes the set of functions. We use $\mathcal C$,
ranged over by $C$, to denote the set of platform configurations. A
configuration associates each of its workers (in \(\Works\)) with a triple
relating the multiset of functions (\(\mathit{Multiset}(\Funcs)\)) currently
allocated on that worker, the amount of resources (in \(\mathbb{N}\)) used by
such functions, and the maximal amount of resources (also in \(\mathbb{N}\))
available to that worker. Functions are tagged to associate them with a
scheduling policy.
We use $\Tags$, ranged over by $t$, to
denote the set of tags and define \reg{} (short for registry) as a map that
associates each function with its tag and its occupancy, i.e., the amount of
resources needed to host it. \(\mathbb{N}\) represent the natural numbers---even
considering fractional resources, we deem naturals enough fine-grained for our
purpose, since we can always convert these to \(\mathbb{N}\) with a constant
multiplying factor.
We use \(app\) to denote the set of \app scripts written following the grammar
presented in \cref{fig:app_syntax} without the \hln{affinity} construct.

\label{sec:encoding}
\begin{figure}[h]
\begin{adjustbox}{width=.9\textwidth}
\begin{minipage}{\textwidth}
{\footnotesize
\[
  \enc{\many{-\ \textit{tag}} \ ::\ -\ \mathtt{default}\hlstr{:} \many{\mathtt{-}\ \textit{block}}}
  = \enc{\many{-\ \textit{tag}} \ ::\ -\ \mathtt{default}\hlstr{:} \many{\mathtt{-}\ \textit{block}}
  \quad \hln{followup} \hlstr{:}\ \hlopt{fail}}
\]
\[
  \begin{array}{rll}
    \enc{\many{-\ \textit{tag}} \ ::\ -\ \mathtt{default}\hlstr{:} \many{\mathtt{-}\ \textit{block}}
    \quad \hln{followup} \hlstr{:}\ \fopt}
    & = & \bigcup\limits_{t \ \in\ \many{\textit{tag}}} \left\{\enc{t}_{\many{\enc{\textit{block}}}}\right\}
  \\
  \enc{\textit{id} \ \hlstr{:} \ \many{\mathtt{-}\ \textit{block} }}_{\many{\block}} =
  \enc{\textit{id} \ \hlstr{:} \ \many{\mathtt{-}\ \textit{block} } \quad
    \hln{followup}\ \hlstr{:}\ \hlopt{default}}_{\many{\block}} & = &
    (\textit{id}, \many{\enc{\textit{block}}} :: \many{\block})
  \\[.5em]
  \enc{\textit{id} \ \hlstr{:} \ \many{\mathtt{-}\ \textit{block} } \quad
    \hln{followup}\ \hlstr{:}\ \hlopt{fail}}_{\many{\block}} & = &
    (\textit{id}, \many{\enc{\textit{block}}})
    \\
    \enc{\hln{workers} \hlstr{:}\ \wopt}
    & = & ( \enc{\textit{\wopt}}, \hlopt{any}, \hlopt{capacity\_used}\ 100\hlopt{\%}::\emptyseq)
    \\
    \enc{\hln{workers} \hlstr{:}\ \wopt
    \quad \hln{strategy} \ \hlstr{:}\ \sopt}
    & = & ( \enc{\wopt}, \sopt, \hlopt{capacity\_used}\ 100\hlopt{\%}::\emptyseq )
    \\
    \enc{\hln{workers} \hlstr{:}\ \wopt
      \quad \hln{invalidate} \ \hlstr{:}\ \many{-\ \iopt}}
    & = & ( \enc{\wopt}, \hlopt{any}, \many{\iopt} )
    \\
    \enc{\hln{workers} \hlstr{:}\ \wopt
      \quad \hln{strategy} \ \hlstr{:}\ \sopt
      \quad \hln{invalidate} \ \hlstr{:}\ \many{-\ \iopt}}
    & = & ( \enc{\wopt}, \sopt, \many{\iopt} )
    \\
    \enc{\hlopt{*}} & = & \hlopt{*}
    \\
    \enc{\many{-\ id}} & = & \many{\ id}
    \\
  \end{array}
\]
}
\end{minipage}
\end{adjustbox}
\caption{\app Syntax Encoding.}
\label{fig:encoder}
\end{figure}

\subsubsection{From YAML to \app{} Semantics' Structures}
In our formal model of {\app}, we need to represent scripts as mathematical
objects. Formally, we define a straightforward encoding $\enc{\cdot}$ that,
given a script in $app$, returns a policy function $\policy$ (ranging over the
set $\App$) with all \hln{followup}s unfolded---where \texttt{default} always
\hlopt{fail}s. The encoding, reported in \cref{fig:encoder}, inductively walks
through the syntax of the \app script and translates each fragment into the
corresponding mathematical object in \(\Policies\).
The only notable bits of the encoding regard the inclusion of the standard
options for the missing parameters---for \hln{strategy} we set it to \hlopt{any}
and for \hln{invalidate} we set it to the maximal capacity of the worker, i.e.,
\hlopt{capacity\_used}\ 100\hlopt{\%}---and the static resolution of the
\hln{followup} parameter, where we concatenate the list of blocks of the tag
with the blocks of the \code{default} one, in case the \hlopt{default} option is
present. Notation-wise, we introduce \(::\) and \(\emptyseq\) to resp.\@ indicate
list concatenation and empty sequence (frequently omitted for brevity).

\subsubsection*{LTS Rules and Examples}
\begin{figure}[!t]
  \begin{small}
    \begin{center}
      \begin{gather*}
        \rulelabel{Workers Layer}
        \\
        \inference[\rName{W_{first}}]
        {\strat(\many{w},s) = w & w \neq \bot &\valid(f,w,\many{i},C)}
        {f,\many{w},s,\many{i},C \triangleright w}
        \qquad
        \inference[\rName{W_{end}}]
        {\strat(\many{w},s)=\bot}
        {f,\many{w},s,\many{i},C \triangleright \bot}
        \\[.3em]
        \inference[\rName{W_{next}}]
        {\strat(\many{w},s) = w & w \neq \bot & \neg \valid(f,w,\many{i},C) & f,\many{w}\setminus w,s,\many{i},C \triangleright w'}
        {f,\many{w},s,\many{i},C \triangleright w'}
        \\[.3em]
        \rulelabel{Blocks Layer}
        \\[.3em]
        \inference[\rName{B_{one}}]
        {\many{w'} = \many{w} \cap \dom(C) & f,\many{w'},s,\many{i},C \triangleright w}
        {f,C,(\many{w},s,\many{i}) \to w}
        \quad
        \inference[\rName{B_{star}}]
        {f,\many{\dom(C)},s,\many{i},C \triangleright w}
        {f,C,(\hlopt{*},s,\many{i}) \to w}
        \\[.3em]
        \inference[\rName{B_{first}}]
        {f,C,\block_1 \to w & w \neq \bot}
        {f,C,\block_1::\cdots::\block_n \to w}
        \qquad
        \inference[\rName{B_{next}}]
        {f,C,\block_1 \to \bot & f,C,\block_2::\cdots::\block_n \to w} 
        {f,C,\block_1::\block_2::\cdots::\block_n \to w}
        \\[.3em]
        \rulelabel{Configuration Layer}
        \\[.3em]
        \inference[\rName{C_{start}}]
        {reg(f) = (n,t)
          & \policy(t) = \many{\block}
          & f,C,\many{\block} \rightarrow w
          & w \neq \bot
          & C( w ) = (\sigma,n',m)
        }
        {C \xRightarrow{(start, f, w)} C[ w \mapsto (\sigma \cup \{f\},n'+n,m)]}
        \\[.2em]
        \inference[\rName{C_{fail}}]
        {
          reg(f) = (\cdot,t)
          & \policy(t) = \many{\block}
          & f,C,\many{\block} \rightarrow \bot
        }
        {C \xRightarrow{(fail,f)} C}
        \\
        \inference[\rName{C_{done}}]
        {
          w \in \dom(C) 
          & f \in \sigma 
          & reg(f) = (n,\cdot)
          & C(w) = (\sigma, n', m) 
        }
        {C \xRightarrow{(done,f,w)} C[ w \mapsto (\sigma \setminus \{f\}, n'-n,m) ]}
        \\
        \rulelabel{\strat{} relation and \valid{} predicate}
        \\
        \strat(\many{w},s) = \begin{cases}
          w    & \mbox{if } \begin{array}[t]{l}
                              s = \mbox{\hlopt{any}}
                              \wedge w \in \{\many{w}\}
                            \end{array}
          \\
          w    & \mbox{if } \begin{array}[t]{l}
                              s = \mbox{\hlopt{best\_first}}
                              \wedge \many{w} = w::\many{w'}
                            \end{array}
          \\
          \bot & \mbox{otherwise}
        \end{cases}
        \\
        \valid(f,w,i_1::\cdots::i_n,C) = \valid(f,w,i_1,C) \ \wedge\ \cdots \ \wedge\ \valid(f,w,i_n,C)
        \\
        \valid(f,w,i,C) = \begin{cases}
          \textsf{true}  & \mbox{if } \begin{array}[t]{l}
                                        i = \mbox{\hlopt{capacity\_used}}\ n \hlopt{\%}
                                        \\ \wedge\ reg( f ) = (n_f, \cdot)
                                           \wedge\ C( w ) = (\cdot, n_{cur}, n_{max})
                                        \\ \wedge\ \min( n, 100 ) \geq 100 * ( n_{cur} + n_f ) / n_{max}
                                      \end{array}
          \\[3em]
          \textsf{true}  & \mbox{if } \begin{array}[t]{l}
                                        i = \mbox{\hlopt{max\_concurrent\_invocations}}\ n
                                        \\ \wedge\ \valid(f,w,\hlopt{capacity\_used}\ 100 \hlopt{\%},C)
                                        \\ \wedge\ C( w ) = (\sigma, \cdot, \cdot)
                                           \wedge\ n \geq |\sigma| + 1
                                      \end{array}
          \\[1em]
          \textsf{false} & \mbox{otherwise}
        \end{cases}
      \end{gather*}
    \end{center}
    \caption{\(\strat\) and \(\valid\) functions.}
    \vspace{-1em}
    \label{fig:rules}
  \end{small}
\end{figure}

We can now present and comment on the rules of the LTS on configurations,
reported in \cref{fig:rules}.
The semantics has three layers: \textit{a)} Configuration
rules (we prefix their names with \(C\)), \textit{b)} Blocks rules (prefixed
with \(B\)), and \textit{c)} Workers rules (prefixed with \(W\)). At the bottom
of \cref{fig:rules}, we define the auxiliary relations \strat{} and \valid{},
used by the Workers rules to check respectively if a worker can be selected
according to a given strategy and if the allocation of the function does not
violate any constraint on the selected worker.

{\em Configuration Rules}
The LTS in \cref{fig:rules} has three kinds of labels, ranged over by
\(\lambda\): $(start,f,w)$ indicates the allocation of an instance of function
$f$ on the worker $w$, $(done,f,w)$ denotes the deallocation on $f$ on $w$, and
$(\mathit{fail},f)$ traces the failure to schedule \(f\) on the current
configuration.
Recalling the \(C_0\) and \(C_1\) presented above, the labelled transition $C_0
\xRightarrow{(start,f,\texttt{w1})} C_1$ represents a reduction
from the configuration $C_0$ to $C_1$ upon the allocation of $f$ on the worker
$\texttt{w1}$.

We focus on the two actions that can change the state of a given configuration:
the allocation of a new function instance (\rName{C_{start}}) and the deallocation of a function instance (\rName{C_{done}})---rule
\rName{C_{fail}} tracks failed scheduling attempts, but has no effects on
configurations. The rules use the function-update notation \(C[ \cdot \mapsto
\cdot ]\) to model the allocation and removal of functions on workers.
Specifically, in \rName{C_{start}} we allocate the function by joining the
multiset \(\sigma\) of allocated functions on the worker \(w\) with the new
function instance \(f\). We also update the current occupancy of the worker with
the units of the function. In \rName{C_{done}}, we remove the function from the
worker by subtracting one instance from \(\sigma\) and removing its units
from the current occupancy of the worker.

\begin{figure}[t]
\begin{lstlisting}[xleftmargin=.3\textwidth, language=yaml, mathescape, basicstyle=\linespread{1}\footnotesize\ttfamily,caption={Example APP script.},label=lst:example_app]
- f_tag:
  - $\hln{workers}$:
      - w1
      - w2
    $\hln{strategy}\hlstr{:}\ \hlopt{best\_first}$
    $\hln{invalidate}$:
      - $\hlopt{capacity\_used}$ 80%
  $\hln{followup}\hlstr{:}\ \hlopt{fail}$
\end{lstlisting}
\end{figure}

To exemplify the rules, we consider the example in \Cref{lst:example_app} of
\app script and three configurations \(C_0\), \(C_1\), \(C_2\). We consider an
infrastructure that includes two workers: \texttt{w1} with maximal capacity 10
and \texttt{w2} with maximal capacity 20 and we assume that \(f\) takes 8 units.
Let \(C_0\) be the configuration where the workers have no functions allocated
on them. To obtain \(C_1\) we start from \(C_0\) and allocate \(f\) on
\code{w1}. \(C_2\) is a reduction from \(C_1\), where we schedule another time
\(f\); since \code{w1} is full (it cannot host \(f\) due to the invalidation
conditions of \code{f_tag}), we allocate \(f\) on \code{w2}. Formally (omitting
\(C_1\) for brevity):

\[
C_0 = \{\
\mathtt{w1} \mapsto (\emptyset,\ 0,\ 10),\
\mathtt{w2} \mapsto (\emptyset,\ 0,\ 20)\ \}
\]
\[
C_2 = \{\
\mathtt{w1} \mapsto (\{f\},\ 8,\ 10),\
\mathtt{w2} \mapsto (\{f\},\ 8,\ 20)\ \}
\]

\begin{figure}[t]
    \begin{subequations}
    \begin{align}
    \parbox{\textwidth}{\adjustbox{width=.99\textwidth}{
    \(\inference[\rName{C_{start}}]
    {\reg(f) = (8,\mcode{f_tag})
      & \policy(\mcode{f_tag}) = \mathit{block}_f::\emptyseq
      & \inference[\rName{B_{first}}]{\cref{ex:e_rule}}{f,C_0,\mathit{block}_f::\emptyseq \rightarrow \mcode{w1}}
      & \mcode{w1} \neq \bot
      & C_0( \mcode{w1} ) = (\emptyset, 0,10)
    }
    {C_0 \xRightarrow{(start, f, \mcode{w1})} C_0[ \mcode{w1} \mapsto (\emptyset \cup \{f\},0+8,10) ]}
    \label{ex:a_rule}
    \)}\vspace*{-8em}}
    \\
    \parbox{\textwidth}{\resizebox{.99\textwidth}{!}{
    \(\inference[\rName{C_{start}}]
    {\reg(f) = (8,\mcode{f_tag})
      & \policy(\mcode{f_tag}) = \mathit{block}_f::\emptyseq
      & \inference[\rName{B_{first}}]{\vdots}{f,C_1,\mathit{block}_f::\emptyseq \rightarrow \mcode{w2}}
      & \mcode{w2} \neq \bot
      & C_1( \mcode{w2} ) = (\emptyset, 0, 20)
    }
    {C_1 \xRightarrow{(start, f, \mcode{w2})} C_1[ \mcode{w2} \mapsto (\emptyset \cup\{f\},0+8,20) ]}
    \label{ex:a_rule_other}
    \)}}
    \end{align}
    \end{subequations}
    \vspace{-1em}
    \end{figure}

We illustrate the rules by considering the allocation of an instance of function $f$ on a worker, where \(\policy\) is the one we obtained from the use case encoding, and the configuration
is $C_0$. The reduction, reported in
\cref{ex:a_rule}, happens via rule \rName{C_{start}}. There, $f$ has tag
\texttt{f\_tag} and $\policy$ maps it to \(\mathit{block}_f\) (cf.
\cref{sec:encoding}). The third premise uses the Blocks rules
(\rName{B_{first}}, derived in \cref{ex:e_rule}) to find a valid (non-\(\bot\))
worker (\texttt{w1}). We update \(C_0\) by allocating \(f\) on \texttt{w1}.
In \cref{ex:a_rule_other}, we want to allocate \(f\) given configuration
\(C_1\). Since \texttt{w1} has insufficient capacity,
it has 8 over 10 units allocated to another instance of \(f\)) the application
of rule \rName{C_{start}} selects the second worker (\(\texttt{w2}\))---obtaining configuration \(C_2\). 

\indent
{\em Blocks Rules} The rules in the blocks layer embody the logic of block
allocation unfolding. Briefly, we pick the first block (\rName{B_{first}}), and
check if we can find an eligible worker for the allocation within that
block---either given a list of workers (\rName{B_{one}}) or the universal
(\rName{B_{star}}).
If that is not the case, we continue by unfolding the list of blocks following
their order of appearance (\rName{B_{next}}). Notice that, in rule
\rName{B_{star}}, we abuse the list notation \(\many{\dom(C)}\) to indicate the
transformation of the domain of \(C\) into a list of workers. We intend this
transformation idempotent, i.e., transformations of the same set always result in the same sequence of elements.

Continuing our analysis of the examples, \cref{ex:e_rule} shows the evaluation
of $\mathit{block}_f$ started in \cref{ex:a_rule}---on the left of the operator,
\(\to\), we replaced \(\mathit{block}_f\) with its components. \textit{Nomen omen}, the rule \rName{B_{first}}
successfully finds a valid worker from the first (and only) block, using the
rule \rName{B_{one}}. In that rule, we find the candidate workers at the
intersection between those provided by \(C_0\) and the ones listed in the block
(\(\texttt{w1}::\texttt{w2}\)). Then, we use the selection rules
(\rName{W_{first}}, derived in \cref{ex:s_rule}) to find a valid worker.
We omit to report the reduction for the second scenario on \(C_1\) since it is
similar to the previous one at blocks level. However, we report in 
\cref{ex:s_rule_other} the upper layers of that reduction to illustrate rule \rName{W_{next}}.\\
%
\begin{figure}[t]
\begin{equation}
\hspace{5em}
\resizebox{.76\textwidth}{!}{
  \(\inference[\rName{B_{first}}]
  {\inference[\rName{B_{one}}]
  {\mcode{w1}::\mcode{w2} = \many{w} \cap \{ \mcode{w1}, \mcode{w2} \} & 
  \inference[\rName{W_{first}}]{\cref{ex:s_rule}}{f,\mcode{w1}::\mcode{w2},s,\many{i},C_0 \triangleright \mcode{w1}}}
  {f,C_0,(\many{w},s,\many{i}) \to \mcode{w1}} & \mcode{w1} \neq \bot}
  {f,C_0,(\ \underbrace{\mcode{w1}::\mcode{w2}}_{\many{w}},\ \underbrace{\hlopt{best\_first}}_{s},\ \underbrace{\hlopt{capacity\_used}\ 80\%}_{\many{i}}\ )::\emptyseq \to \mcode{w1}}
  \label{ex:e_rule} 
\)}
\end{equation}
\begin{subequations}
\begin{align}
  \resizebox{.72\textwidth}{!}{
    \(\inference[\rName{W_{first}}]
    {\strat(\mcode{w1}::\mcode{w2},\hlopt{best\_first}) = \mcode{w1} & \mcode{w1} \neq \bot &\valid(f,\mcode{w1},\many{i},C_0)}
    {f,\mcode{w1}::\mcode{w2},\hlopt{best\_first}, \underbrace{\hlopt{capacity\_used}\ 80\%}_{\many{i}},C_0 \triangleright \mcode{w1}}
    \label{ex:s_rule} 
  \)}
  \\
  \resizebox{.88\textwidth}{!}{
    \(\inference[\rName{W_{next}}]{
    \strat(\many{w},s) = \mcode{w1}
    &
    \mcode{w1} \neq \bot & \neg \valid(f,\mcode{w1},\many{i},C_1) 
    &
    \inference[\rName{W_{first}}]{
      \vdots
    }{
      f,\many{w} \setminus \mcode{w1},s,\many{i},C_1 \triangleright \mcode{w2}
    }
    }{
      f,\underbrace{\mcode{w1}::\mcode{w2}}_{\many{w}},\underbrace{\hlopt{best\_first}}_{s}, \underbrace{\hlopt{capacity\_used}\ 80\%}_{\many{i}},C_1 
      \triangleright \mcode{w2}
    }
    \label{ex:s_rule_other} 
  \)}
\end{align}
\end{subequations}
\end{figure}
%
%
%
\indent
{\em Workers Rules} The last stratum of the \app LTS semantics are the
Workers rules, which evaluate strategies and invalidation policies to find (if
any) a valid worker to allocate the function. The logic of this layer
is to exhaustively try to find a valid worker among the ones listed (in a
block). Specifically, we either find the first worker \valid{} (selected
following the strategy \(s\) of the block) and use that one (\rName{W_{first}})
or we cycle through the remaining workers in the list (\rName{W_{next}}), either
returning the first \valid{} worker or exhausting all possibilities, reporting
failure (\rName{W_{end}}).

Completing our examples, \cref{ex:s_rule} ends the derivation from
\cref{ex:a_rule} and \cref{ex:e_rule} via rule \rName{W_{first}}. There, the
\strat{} relation (left-most premise) interprets the \hln{strategy} of the block
and finds a worker (\code{w1}) while the \valid{} predicate (right-most premise)
checks if the worker can support the allocation of the function---e.g.,
considering the list of \hln{invalidate} options of the block.

\cref{ex:s_rule_other} closes the derivation on \(C_1\) started with
\cref{ex:a_rule_other}, and shows the logic of workers selection with invalid
workers. Since \code{w1} lacks the capacity to host \(f\) (\(\neg
\valid({\cdots})\)), in the right-most premise, we remove it from the candidate
workers and find \code{w2} with rule \rName{W_{first}}.

\subsection{\appp Semantics}

To define the semantics of \appp, we need to extend the mathematical objects of the formal model of \app{} to keep track of the (anti-)affinity constraints. The extension regards the blocks \(\Blocks\), which, from triples, become quintuples, where the last two elements are respectively the affine and anti-affine list of tags of the block.
\[
  \begin{array}{c}
  \Blocks \triangleq (\mathit{List}(\Works) \cup \hlopt{*}) \times \textit{\sopt} \times \mathit{List}(\textit{\iopt}) \times \mathit{Identifiers} \times \mathit{Identifiers}
  \end{array}
\]

Due to the extension of blocks, we need also to define the encoding function \(\encp{\cdot}\) to translate \appp{} scripts into the extended mathematical representation of policy functions. In \cref{fig:encoder_appp}, we report the encoding function \(\encp{\cdot}\) from
the syntax to the objects used in the operational semantics. The interesting bit in the definition of \(\encp{\cdot}\) is the revision of the block-encoding rule (third from the bottom). The rule reuses the related rule in \cref{fig:encoder} (\(\enc{\cdot}\)) to encode the fragments already present in \app and it hands the encoding of the new \hln{affinity} clause to the rules \(\encp{\cdot}_{+}\) and \(\encp{\cdot}_{-}\) resp.\ for the affine and anti-affine tags.

\begin{figure}[h]
\begin{adjustbox}{width=.76\textwidth}
\begin{minipage}{\textwidth}

\[
  \encp{\many{-\ \textit{tag}} \ ::\ -\ \textit{default}\ \hlstr{:} \many{\mathtt{-}\ \textit{block}}}
  = \encp{\many{-\ \textit{tag}} \ ::\ -\ \textit{default}\ \hlstr{:} \many{\mathtt{-}\ \textit{block}}
  \quad \hln{followup} \hlstr{:}\ \hlopt{fail}}
\]
\[
  \begin{array}{rll}
    \encp{\many{-\ \textit{tag}} \ ::\ -\ \textit{default}\ \hlstr{:} \many{\mathtt{-}\ \textit{block}}
    \quad \hln{followup} \hlstr{:}\ \fopt}
    & = & \bigcup\limits_{t \ \in\ \many{\textit{tag}}} \left\{\encp{t}_{\many{\encp{\textit{block}}}}\right\}
  \\
  \encp{\textit{id} \ \hlstr{:} \ \many{\mathtt{-}\ \textit{block} }}_{\many{\beta}} =
  \encp{\textit{id} \ \hlstr{:} \ \many{\mathtt{-}\ \textit{block} } \quad
    \hln{followup}\ \hlstr{:}\ \hlopt{default}}_{\many{\beta}} & = &
    (\textit{id}, \many{\encp{\textit{block}}} :: \many{\beta})
  \\[.5em]
  \encp{\textit{id} \ \hlstr{:} \ \many{\mathtt{-}\ \textit{block} } \quad
    \hln{followup}\ \hlstr{:}\ \hlopt{fail}}_{\many{\beta}} & = &
    (\textit{id}, \many{\encp{\textit{block}}})
  \end{array}
\]
\[
  \begin{array}{rll}
    \encp{\textit{block}} & = & \begin{cases}
      (\wopt,\sopt,\many{\iopt},\emptyseq,\emptyseq) &
        \mbox{if } \enc{\textit{block}} = (\wopt,\sopt,\many{\iopt})
      \\[1em]
      (\wopt,\sopt,\many{\iopt},\encp{\many{\textit{\aopt}}}_{+}, \encp{\many{\textit{\aopt}}}_{-}) &
        \mbox{otherwise, let } \textit{block} = \mbox{``}\bullet \quad \hln{affinity} \ \hlstr{:}\ \many{-\ \textit{\aopt}}\ \mbox{''} \\ & \mbox{and} \ \enc{\bullet} = (\wopt,\sopt,\many{\iopt})
    \end{cases}
    \\
      	\encp{\emptyseq}_{\bullet} & = & \emptyseq
	\\
    \encp{\aopt :: \many{\aopt}}_{\bullet} & = &
    \begin{cases}
      id :: \encp{\many{\aopt}}_{\bullet} &
      \mbox{if } ( \aopt = \textit{id} \ \wedge \ \bullet = + )
      \vee ( \aopt = \textit{!id} \ \wedge \ \bullet = - )
    \\
    \encp{\many{\aopt}}_{\bullet} & \mbox{otherwise}
  \end{cases}
  \end{array}
\]
\end{minipage}
\end{adjustbox}
\caption{\appp
encoding.}
\vspace{-1em}
\label{fig:encoder_appp}
\end{figure}

\subsubsection{LTS rules}

In \cref{fig:rules_app_plus}, we report the definition of the operational
semantics of {\appp}.
In particular, we update the rules \rName{B_{one}} and
\rName{B_{star}} of the semantics of \app, to adapt them to the shape of the new
blocks, which carry additional information on affine and anti-affine tags.
The interesting addition lies in the usage, in the Workers rules, of an
overloaded version of the \valid{} predicate that verifies if we can allocate
the function, checking both the capacity constraints of the block and the
presence (resp. absence) of the (functions associated with) tags that are affine
(resp. anti-affine) to the block. Formally, the 6-parameter \valid{} predicate
holds if \emph{a}) the capacity constraints hold (i.e., if the 4-parameter
\valid{} predicate holds, cf. \cref{fig:rules}) and \emph{b}) all affine tags
are present in \(w\), \(\{\many{id}\}\setminus t = \emptyset\), and all
anti-affine tags are absent from \(w\), \(\{\many{id'}\}\cap t = \emptyset\) ---
where \(t\) is the set of all the tags of the functions allocated on the
selected worker \(w\), found in the associated multiset \(\sigma\), and
\(\many{id}\) and \(\many{id'}\) are resp.\ the list of affine and anti-affine
tags of the block.

\begin{figure}[t]
  \begin{adjustbox}{width=.9\textwidth}
  \begin{minipage}{\textwidth}
    \begin{center}
      \begin{gather*}
        \rulelabel{Workers Layer}
        \\
        \inference[\rName{W_{first}}]
        {\strat(\many{w},s) = w & w \neq \bot &\valid(f,w,\many{i},\many{id},\many{id'},C)}
        {f,\many{w},s,\many{i},\many{id},\many{id'},C \triangleright w}
        \qquad
        \inference[\rName{W_{end}}]
        {\strat(\many{w},s)=\bot}
        {f,\many{w},s,\many{i},\many{id},\many{id'},C \triangleright \bot}
        \\[.3em]
        \inference[\rName{W_{next}}]
        {\strat(\many{w},s) = w & w \neq \bot & \neg \valid(f,w,\many{i},\many{id},\many{id'},C) & f,\many{w}\setminus w,s,\many{i},C \triangleright w'}
        {f,\many{w},s,\many{i},\many{id},\many{id'},C \triangleright w'}
        \\[.3em]
        \rulelabel{Blocks Layer}
        \\[.3em]
        \inference[\rName{B_{one}}]
        {\many{w'} = \many{w} \cap \dom(C) & f,\many{w'},s,\many{i},\many{id},\many{id'},C \triangleright w}
        {f,C,(\many{w},s,\many{i},\many{id},\many{id'}) \to w}
        \quad
        \inference[\rName{B_{star}}]
        {f,\many{\dom(C)},s,\many{i},\many{id},\many{id'},C \triangleright w}
        {f,C,(\hlopt{*},s,\many{i},\many{id}, \many{id'}) \to w}
        \\[.3em]
        \rulelabel{\valid{} predicate (overloaded)}
        \\[.3em]
        \valid(f,w,\many{i},\many{id}, \many{id'},C) = \valid(f,w,\many{i},C)
        \ \wedge\ \{\many{id}\}\setminus t = \{\many{id'}\}\cap t = \emptyset \\
        \mbox{where } \ C(w) = (\sigma, \cdot,\cdot) \mbox{ and }
        t = \{ id\ |\ f \in \sigma \ \wedge\ reg(f) = (\cdot, id)  \}
      \end{gather*}
    \end{center}
    \end{minipage}
    \end{adjustbox}
    \caption{\appp{} Semantics — uses the rules from \cref{fig:rules} except for the one overwritten here.}
    \vspace{-1em}
    \label{fig:rules_app_plus}
\end{figure}

\section{\cref{theo:PSPACE_pp}}
\label{appendix-theorems}

In this section, we report the full proof of \cref{theo:PSPACE_pp} that for space reason has been only sketched
in the main part of the paper.

\setcounter{theorem}{3}
\begin{theorem} In positively-polarised {\appp}, i.e.,
    where one can only express affinity constraints, the problems
    \Reach($p,reg,C,f,w$) and \CoOccur($p, reg, C,\{f,g\},w$) are NP-hard.
    \end{theorem}
    \vspace{-5mm}
    
    \begin{proof} 
    We start by proving that the problem \Reach($p,reg,C,f,w$) is NP-hard. This
    result is proved by reduction from 3SAT, a well known NP-hard problem
    \cite{DBLP:conf/stoc/Cook71} consisting of checking the satisfiability of a
    boolean formula in conjunctive normal form, where each clause has at most three
    litarals (where a literal is a boolean variable or its negation).
    
    Consider a boolean formula $\varphi$ with $n$ variables $x_1, x_2,\dots,x_n$ and
    $m$ clauses $c_1,c_2,\dots,c_m$ each one with literals $l_j^1,l_j^2,l_j^3$. The
    idea behind the reduction is to encode each possible literal with a
    corresponding function $l_i$ that can be scheduled. We use the capacity
    constraints to limit the possibility to schedule both $l_i$ and its negation
    $!l_i$. Each clause is encoded with a corresponding function $c_j$ that can
    be scheduled only if at least one of its literal functions $l_j^1,l_j^2,l_j^3$
    has been already scheduled. We have that all such functions $c_j$
    can be scheduled if and only if the formula $\varphi$ is satisfiable.
    
    We consider a unique worker $w$ with capacity $3*2^{n+2}-2$.
    
    For each variable $x_i$ we consider five functions $x_i$, $!x_i$, $left_i$,
    $mid_i$, $right_i$ all consuming $2^i$ resources, and with $x_i$, $!x_i$
    having max capacity $(4 * 2^i-1) * 100 / (3*2^{n+2}-2)$. Moreover, we impose
    $x_i$ affine to $left_i$ and $mid_i$, and $!x_i$ affine to $mid_i$ and
    $right_i$.
    
    Note that with such construction, it is not possible to have the functions $x_i$
    and $!x_i$ both scheduled on the worker $w$ because this requires to
    schedule at the same time $x_i$,  $!x_i$, $mid_i$ and one between $left_i$
    and $right_i$, violating the capacity constraint.
    
    For each clause, we consider one function $c_j$ consuming $2^{n+2}$ resources.
    Each function $c_j$ has three scheduling blocks. For the first function $c_1$
    the first block imposes $l_1^1$ as an affine function, the second one considers
    $l_1^2$, while the third one $l_1^3$. For $j>1$, each scheduling block imposes
    an affinity with also the function $c_{j-1}$, besides the corresponding
    $l_j^1,l_j^2,l_j^3$ literal functions, respectively.
    Note that when a function $c_j$ is scheduled, it is not possible to schedule any
    $x_i$ or $!x_i$: the size of $c_j$ is $2^{n+2}$ that is greater than the
    biggest capacity constraint for $x_i$ (i.e., no more than $4 * 2^n-1$).
    Moreover, note that the encoding is polynomial in the size of the input since
    representing the formula would take space $O(n+m)$ and representing $2^n$ can be
    done in space $O(n)$.
    
    We have that $\varphi$ is satisfiable if and only if $c_m$ can be scheduled on
    $w$ (assuming $w$ empty in the initial configuration).
    The left-to-right implication follows from the existence of a sequence in which
    the literals which satisfy $\varphi$ are scheduled on $w$ following the order
    implied by the indexes of the variables: either $x_1$ or $!x_1$ at the
    beginning, then $x_2$ or $!x_2$, etc.
    Following this scheduling order is possible since the functions for the first
    $k$ literals take space $\sum_{i=1}^k 2^i = 2^{k+1}-2$, and no additional space
    is taken from the auxiliary functions $left_i$, $mid_i$, $right_i$ (for $i \leq
    k$) if they are assumed to be removed from $w$ as soon as they are used to
    schedule the corresponding $i$-th literal. This occupation of $w$ does not limit
    the possibility to schedule also the following $k+1$ literal. When the literals
    are all scheduled, the functions $c_1, c_2,\dots,c_m$ can be scheduled in
    sequence; this is possible if we remove the function $c_{j}$ as soon as it has
    been already used to satisfy the affinity of $c_{j+1}$.
    
    The right-to-left implication follows from the observations that it is not
    possible to schedule both a function representing a literal and its negation,
    and the fact that as soon a function $c_j$ is scheduled, no new literals can be
    allocated unless all the functions $c_j$ are removed from $w$. Due to the
    affinity of $c_{j+1}$ for $c_{j}$
    the only possibility to schedule $c_m$ is to have a sequence of scheduling
    actions during which all the functions $c_j$ are scheduled in the expected order
    and there is always at least one of such $c_j$ scheduled on the worker $w$. For
    the above properties, this sequence of scheduling actions cannot contain the
    scheduling of $x_i$ or $!x_i$ functions. Hence, the functions representing
    literals which are present on the worker $w$ when the first $c_1$ function is
    scheduled implies the existence of 
    one assignment of boolean values to the corresponding variables that satisfies
    all the clauses: for each function $x_i$, the corresponding variable should be
    true, for each function $!x_i$, the corresponding variable should be false.
    
    The above reduction from 3SAT proves that \Reach($p,reg,C,f,w$) is NP-hard.
    Using the same arguments in the proof of \Cref{cor:co_occur_PSPACE}, we
    conclude that \CoOccur($p, reg, C,\{f,g\},w$) is not computationally simpler,
    hence it is 
    NP-hard.
    \qed
    \end{proof}

\end{document}


\title{Affinity-aware Serverless Function Scheduling: Semantics, Properties, Analysis, and Implementation}

\section{\appp Implementation}
\label{sec:implementation}

We have implemented and validated an \appp-based
FaaS platform, obtained by extending the \app prototype of Apache
OpenWhisk. Apache OpenWhisk is an open-source, serverless platform
initially developed by IBM and donated to the Apache Software Foundation.
The platform offers a rich programming model for creating serverless APIs from functions, composing functions into serverless workflows, and connecting events to functions using rules and triggers.

The main intervention we performed to make the existing \app{}-based
OpenWhisk architecture \appp{}-compliant consists in an extension of the
\textit{Controller} component, which in the OpenWhisk architecture is
responsible for scheduling functions on the \textit{Invoker} components---the
workers, in OpenWhisk's lingo.
The extension adds a parser for the \appp{} scripts and a new scheduler that
handles the given policies, but the major challenge of implementing \appp{} has
been changing the Apache OpenWhisk's load balancer---the part of the Controller
responsible for scheduling the functions---so that it keeps track of the
functions allocated to all the workers.
We introduced two lookup tables to implement this tracking functionality: the
\textit{activeFunctions} and the \textit{activeTagActivations}.
The first table associates the allocated functions (and their
tags) to their host worker and allows the load balancer to verify affinity and
anti-affinity constraints. The second table is an auxiliary
one. Indeed, to update the \textit{activeFunctions} table, we need to keep track
of the state of the different function instances (possibly of the same function
definition, so we cannot use their identifiers) by pairing their
\emph{activation} \emph{id}s with their function identifiers; when we observe
the termination of an active function, we look its function identifier up and
remove that instance from the \textit{activeFunctions} table---we detect
instance terminations thanks to the messages workers send to notify the load
balancer of their completion.

%
Regarding the affinity and anti-affinity semantics, we consider a function
allocated on a worker from the moment the load balancer sends the allocation
instruction to the selected worker; we consider a function deallocated as soon
as the completion message reaches the load balancer.

We developed the code to implement the update of the functions and changed the
scheduling algorithm in Scala, on a fork of the OpenWhisk
repository~\cite{repository}.
The entire system is easily deployable using
Terraform and Ansible scripts from the dedicated
repository~\cite{repository_deployment}.

To validate our platform,
we use the scenario presented
in \cref{sec:intro} (cf. \cref{fig:example}) to verify that, by enforcing
(anti-)affinity constraints, we can reduce average execution times and tail
latency.
Recalling the example, 
we develop two functions,
\(d\) and \(i\), that represent a simple \emph{divide-et-impera} serverless
architecture running in a realistic co-tenancy context. 
Users invoke \emph{divide} functions, requesting the solution of a
problem. At invocation, \emph{divide} splits the problem into sub-problems and
invokes instances of the second function, \emph{impera}. The \emph{impera}
instances solve their relative sub-problems and store their solution fragments
on a persistent storage service. After the \emph{impera}s terminated,
\emph{divide} retrieves the partial solutions, assembles them, and returns the
response to the user. Referencing the \appp script on the right of
\cref{fig:example}, we indicate \(i\) affine with \(d\).
In the multi-zone execution context of the use case, we have workers from two
data centres (which represent the \(\mathit{Zone}\)s of \cref{fig:example}),
placed far apart from each other. We have two synchronised instances of
persistent storage (like \(db\) and \(db'\) in \cref{fig:example}), one per data
centre. The storage implements an eventual consistency model, i.e., it trades
high availability of data off of its overall consistency.
To minimise latency, both the \emph{divide} and the \emph{impera} functions
access the storage instance closest to them.
Since it can take some time for the two database instances to converge, the
functions implement a traditional exponential back-off retry system---each
function tries to fetch its data (sub-problems/solutions) from its local storage
instance; if the data is not there, starting from a 1-second delay, the function
waits for a back-off time that exponentially increases at each retry.
We also draw the \emph{heavy} functions from \cref{fig:example}, which simulate
the possible interferences of serverless co-tenancy. 

We consider three \app{}/\appp{} scripts to showcase the benefits of
(anti-)affinity constraints.
The first, which uses the full expressiveness of \appp{}, is the one reported on
the right of \cref{fig:example}---where \emph{impera}s are affine with
\emph{divide} and they are both anti-affine with the \emph{heavy} functions.
The second script uses negatively-polarised \appp{} (cf. \cref{sec:appp_other})
and removes the affinity constraints between \emph{impera} and \emph{divide}
from the first script.
The third script omits the anti-affinity constraints from the second one,
effectively making it an \app{} script.

To run the use case, we deploy the OpenWhisk versions of \app{} and \appp{} on a
8-node Kubernetes cluster on the Digital Ocean platform; one node acts as the
control plane (and as such, it is unavailable to OpenWhisk), one hosts the
OpenWhisk core components (i.e., the Controller, the OpenWhisk internal database
CouchDB, and the messaging system 
Kafka), 
and six nodes are workers. We deploy the control plane and the OpenWhisk core
components on virtual machines with 2 vCPU and 2 GB RAM, while we deploy 4
workers on virtual machines with 2 vCPU and 2 GB RAM and 2 workers with 1 vCPU
and 1 GB RAM. All machines run the Ubuntu Server 20.04 OS. Location-wise, we
place the control plane, the OpenWhisk core components, and 3 workers in Europe
and 3 workers in North America (2 with the more powerful configuration and 1
with the lesser one in each zone). To implement persistent storage, we deploy a
2-node MongoDB replica set, one in Europe and one in North America, using the
6.0.2 version of the Community Server.
We distribute the load generated by the \emph{heavy} functions on the platform
with two variants, \textit{heavy\_eu} and \textit{heavy\_us} which, in the
\app{}/\appp{} scripts, we constrain to be resp.\ allocated in the Europe and
the North America data centres on the less powerful workers, to further amplify
the effect of co-tenancy they exert.
%
%
%
All functions are in JavaScript and run on OpenWhisk NodeJS runtime
\textit{nodejs:14}.
\begin{figure}[t]
    \begin{center}
        \includegraphics[width=.99\textwidth]{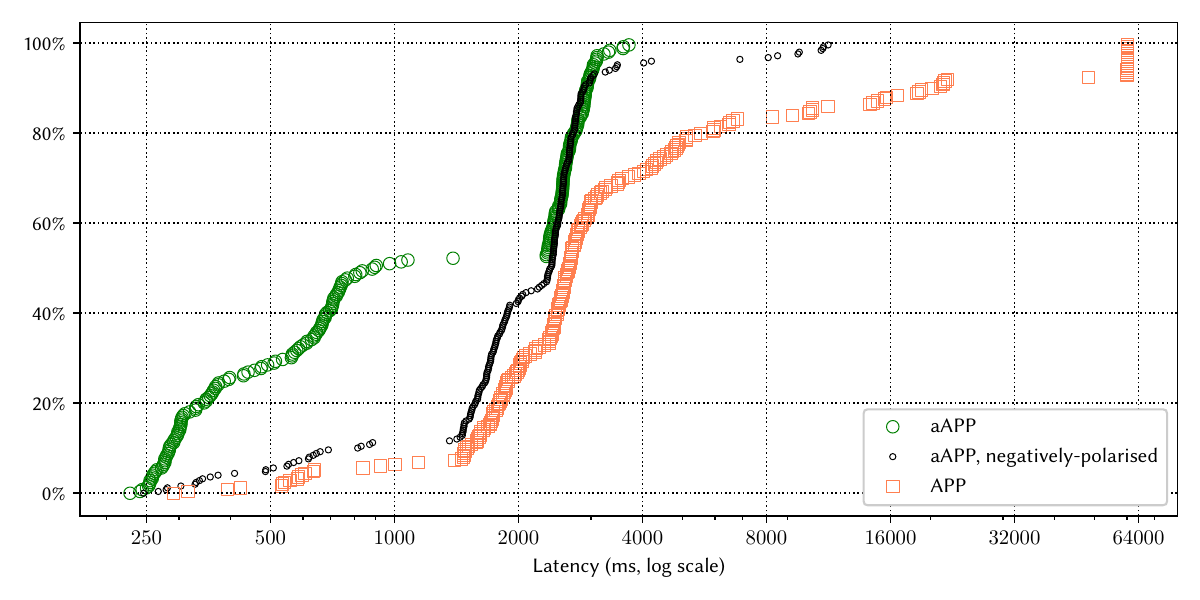}
    \end{center}
    \vspace{-1em}
    \caption{Sorted scatter plot of \textit{divide} functions; \(x\) is the latency (ms) of the $y^\text{th}\%$ fastest invocation.}\vspace{-1em}
    \label{fig:benchmark}
\end{figure}
\vspace{5pt}
\smallpar{Experiments and Results}%
Each experiment involves 5 sequential runs. 
Each run invokes the \emph{heavy\_eu} and \emph{heavy\_us} functions in
non-blocking mode, 
followed by 10 calls of the \textit{divide} function, each one
waiting for the previous to complete. Upon termination of the \emph{heavy}
functions, we proceed with the remaining runs; for a total of 10 \emph{heavy}
and 50 \textit{divide} functions 
per experiment.
To ensure reliable results, we ran the experiment 5 times, totalling 250 calls
of the \textit{divide} function for each of the three \app{}/\appp{} script.
We use Apache JMeter to simulate each request, tracking its latency, number of
retries (to retrieve storage data), and outcomes (success or failure).

The results match our expectations. The mean and median latency for the divide functions in \appp{} is resp.\ 1547ms and 883ms, while
the 95\(^{th}\) tail latency is 3041ms. The corresponding figures increase for neg.\ polarised \appp{}:
2337ms (+40\%), 2381ms (+91\%), and 3476ms (+13\%). As expected, the latency increases even more substantially for \app, with respectively (percentage increase vs \appp{}) 8118ms (+135\%), 2648ms (+99\%), and 60157ms (+180\%).

To further analyse the differences, in \cref{fig:benchmark} we report the
scatter plots where we sort the latencies of the \textit{divide} functions from
the shortest to the longest (\(x\)-axis). We focus on this measure because it offers a comprehensive
overview of the performance of the architecture. In particular, it includes the latencies of the related \textit{impera}
functions and its latencies are concretely the ones experienced by the users interacting with the
system.
%
%
The first striking observation is that the distribution of the \appp data points
is interrupted (there are almost no instances) between the 1000ms and the 2400ms
mark. We attribute this behaviour to having OpenWhisk core components installed
in one region, which exert some overhead on the workers of the other region when
they interact with the platform e.g., to fetch functions and receive/send
requests/notifications. We see similar intervals, although less apparent, for neg.\ polarised \appp and \app.

In the 200--1000ms interval \appp provides consistent, fast
performance, while \app and neg.\ polarised \appp show only a few
well-performing cases---the rest, on the same performance bracket, are shifted
to the right, achieving slower results. We can characterise the ``fast''
invocations as those where the \textit{divide} and its two \textit{impera}
functions appear on a ``free'' node, i.e., without the \textit{heavy} function,
in Europe. Specifically, when using \app, each invocation has a
$\nicefrac{2}{6}$ probability of appearing on a free node in Europe, i.e., the
probability of fast invocations is $(\nicefrac{2}{6})^3 \approx 3.7\%$; using
neg.\ polarised \appp the figure becomes $(\nicefrac{1}{2})^3 = 12.5\%$ (each
invocation has a $\nicefrac{1}{2}$ chance of appearing on a European free node).
Finally, using \appp the probability raises to 50\%, as all three functions go
on the same node (either in the US or in the EU).

Overall, going from \app to neg.\ polarised \appp already improves performance
(mean, median, tail latency improve resp.\ of 110\%, 10\%, and 178\%), which
shows the impact of sharing a worker with \textit{heavy} functions---\app shows a long tail of invocations after the ca. 3000ms mark in
\cref{fig:benchmark}. Looking at worst cases, using \appp does not result in a
considerable performance increase. This is visible from the plot by noticing how
the tail high-percentage instances of neg. polarised \appp and \appp almost
overlap, resulting in a small (+13\%) improvement in tail latency. The
differences in mean (+40\%) and median (+91\%) latency between neg.\ polarised
\appp and \appp emerges in the 250--1000ms bracket, where the former has only a
few data points w.r.t.\@ to the higher number of fast instances of \appp.
Practically, the figures and distribution show how strongly North American
allocations impact latency vs the benefit of co-location. Besides increasing
performance, \appp succeeds in eliminating retries, contrarily to neg.\
polarised \appp{}.

In addition to the above test, we also ran benchmarks to quantify the impact of
the added functionalities of our \appp-based prototype (e.g., the maps that
track the state of functions on workers).
We compared our \appp prototype against both vanilla OpenWhisk and its
\app-based variant under 7 microbenchmarks~\cite{PGMTZ23}, all of which not
bound by affinity.
The general result is that the overhead of supporting \appp-based affinity is
negligible.
Quantitatively, on average, all platforms allocate functions in less than 2ms,
except for a microbenchmark with a big-payload function which takes less than
12ms. As expected, OpenWhisk vanilla is the fastest, closely (under one
millisecond) followed by \app and \appp---except for the big-payload function
microbenchmark, where \app and \appp surprisingly perform better than OpenWhisk.
The differences between \app and \appp are even smaller, with \app being
generally slightly (sub-millisecond) faster than \appp. We refer the interested
reader to \cref{sec:benchmark} for the detailed description of the
microbenchmarks, the configuration settings, and the results of the experiments.

\section*{Comparison between OpenWhisk, \app, and \appp}
\label{sec:experiments_overhead}

In this supplementary material, we provide more details on the experiments
conducted to compare OpenWhisk, its \app-based variant, and our \appp-based
prototype. We start by giving a presentation of the benchmark cases used,
followed by the results related to the latency of the function invocations.

For the cases, we decided to use the benchmark suite used in~\cite{PGMTZ23}.
Considering that in our settings we are not interested in the data locality
capabilities of \app but only in checking the scheduling performances of \appp,
we decided to deploy the platforms in only one cloud zone and use for each
scenario 2000 invocations with the primary goal of simplifying as much as possible
the testing environment and have enough invocation to draw meaningful
comparisons despite the possible variability of the cases.

The benchmarks used are:

\begin{itemize}

    \item \textit{hello-world} (called \textbf{hellojs} in~\cite{PGMTZ23}). It
          implements a ``Hello World'' echo application, and provides an
          indication of the overall performance of the platform when using a
          simple function.

    \item \textit{long-running} (corresponds to \textbf{sleep}
          in~\cite{PGMTZ23}). This function waits 3 seconds, to benchmark the
          handling of multiple functions running for several seconds and the
          management of their queueing process;

    \item \textit{compute-intensive} (corresponds to \textbf{matrixMult}
          in~\cite{PGMTZ23}). This function multiplies two 100x100 matrices and
          returns the result to the caller. This case measures the performance
          of handling functions performing some meaningful computation, and the
          handling of large invocation payloads.

    \item \textit{DB-access (light)} (corresponds to \textbf{mongoDB}
          in~\cite{PGMTZ23}). This function executes a query requiring a
          document from a remote MongoDB database. The requested document is
          lightweight, corresponding to a JSON document of 106 bytes, with
          little impact on computation. This case was used by De Palma et al. to
          measure the impact of data locality on the overall latency. Since all
          workers are from the same cloud zone, this case just measures the
          overhead of scheduling functions that fetch small pieces of data from
          a local database. 

    \item \textit{DB-access (heavy)} (corresponds to \textbf{data-locality}
          in~\cite{PGMTZ23}). This case regards both a memory- and
          bandwidth-heavy data-query function. The function fetches a large
          document (124.38 MB) from a MongoDB database and extracts a property
          from the returned JSON. Similarly to \textit{mongoDB}, in our
          settings, this case only evaluates the overhead of scheduling
          functions that fetch large pieces of data from a local database.

    \item \textit{External service} (corresponds to \textbf{slackpost}
          in~\cite{PGMTZ23}). The case consists of a function that sends a
          message through the Slack API. De Palma et al. draw the function of
          this case from the Wonderless dataset~\cite{Eskandani-S:Wonderless}.

    \item \textit{Code dependencies} (corresponds to \textbf{pycatj}
          in~\cite{PGMTZ23}). The case consists of a formatter that takes an
          incoming JSON string and returns a plain-text one, where key-value
          pairings are translated into Python-compatible dictionary assignments.
          Like the previous case, De Palma et al. draw the function of
          this case from the Wonderless dataset~\cite{Eskandani-S:Wonderless}.

\end{itemize}

Compared to De Palma et al.~\cite{PGMTZ23}, we exclude the \textbf{terrain} and
the \textbf{cold-start} test cases. We excluded the first due to its high-error
rates (De Palma et al. ran the benchmark case but decided to discard those
results due to the high error rate of the case). The second case is redundant
for testing the overhead of scheduling since we capture the features of the case
already with the \textit{hello-world} and \textit{code-dependency} cases. In
particular, the \textbf{cold-start} case is an echo application with sizable,
unused dependencies. The peculiarity of the case is its 10-minute invocation
pattern, used to check the performance of the platform against cold-start times
(so that the platform evicts cached copies of the function, requiring costly
fetch-and-startup times at any subsequent invocation). Since we obtained the same
results with the mentioned, included cases, we discard the much-slower
\textbf{cold-start}.
%

\subsubsection*{Results}

\begin{table}
\begin{adjustbox}{width=\textwidth}
\begin{tabular}{|l|c|c|c|c|c|c|c|c|c|c|c|c|}
\hline
& \multicolumn{4}{c|}{OpenWhisk}
& \multicolumn{4}{c|}{APP}
& \multicolumn{4}{c|}{aAPP} \\
\cline{2-13}
& avg & med & tail lat & st dev & avg & med & tail lat & st dev & avg & med & tail lat & st dev \\ \hline
Hello World & 88 & 84 & 126 & 28 & 78 & 73 & 114 & 23 & 76 & 72 & 109 & 34 \\  
\hline
Long-running & 3118 & 3096 & 3176 & 87 & 3094 & 3074 & 3174 & 86 & 3092 & 3072 & 3175 & 88 \\  
\hline
Compute-intensive & 348 & 330 & 559 & 136 & 304 & 286 & 501 & 122 & 257 & 235 & 409 & 103 \\  
\hline
DB-access (light) & 131 & 111 & 181 & 289 & 119 & 102 & 156 & 241 & 125 & 91 & 139 & 484 \\  
\hline
DB-access (heavy) & 95 & 83 & 130 & 135 & 95 & 84 & 131 & 136 & 87 & 75 & 113 & 158 \\  
\hline
External service & 627 & 613 & 741 & 230 & 640 & 625 & 765 & 308 & 647 & 630 & 778 & 305 \\  
\hline
Code dependencies & 132 & 116 & 213 & 127 & 143 & 117 & 255 & 186 & 98 & 80 & 142 & 209 \\  
\hline
\end{tabular}
\end{adjustbox}
\caption{\label{tab:latencies}Latencies of the benchmarks.}
\end{table}

In \cref{tab:latencies}, we report the latencies of execution of the cases,
characterised by their average (avg), median (med), 95\(^{th}\%\) tail latency
(tail lat), and standard deviation, for each of the three considered platforms. 

\begin{figure}[h]
    \raisebox{-3em}{\includegraphics[width=.33\textwidth]{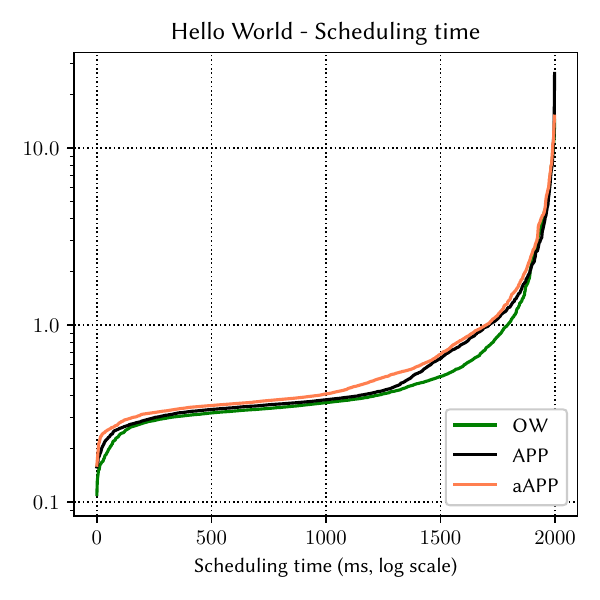}}
    \raisebox{-3em}{\includegraphics[width=.33\textwidth]{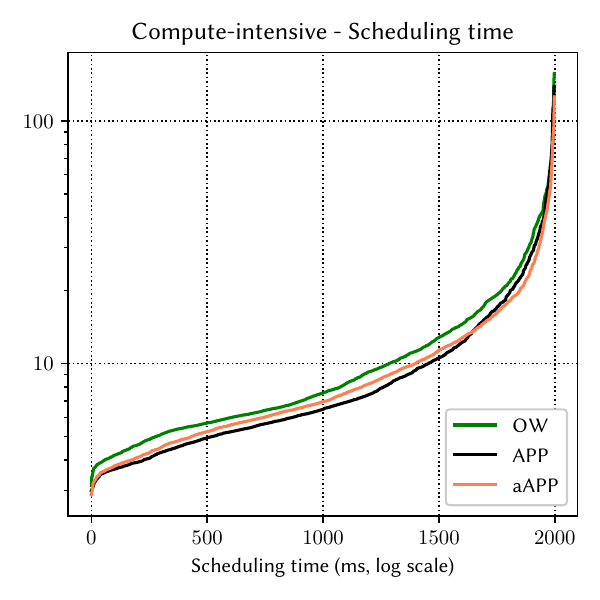}}
    \raisebox{-3em}{\includegraphics[width=.33\textwidth]{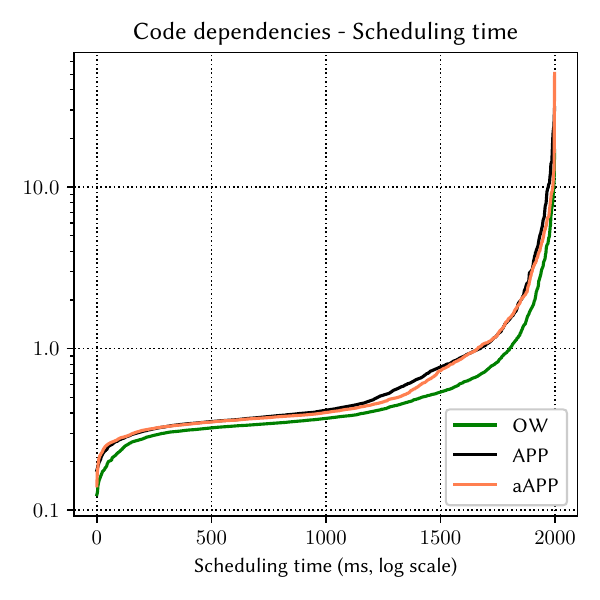}}
    \raisebox{-3em}{\includegraphics[width=.33\textwidth]{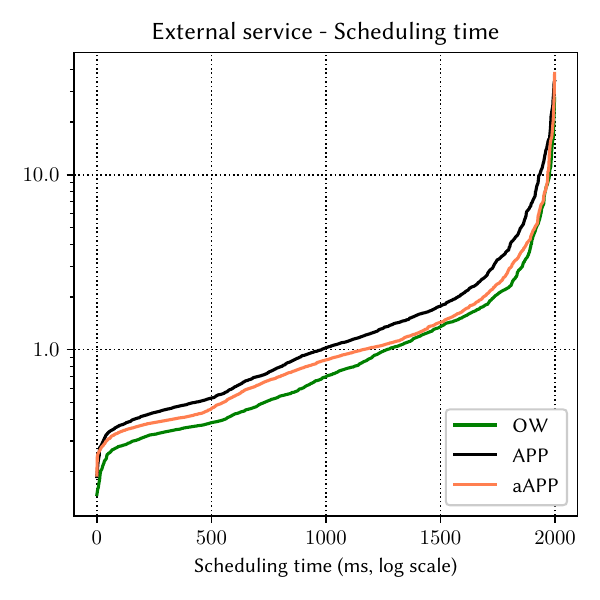}}
    \raisebox{-3em}{\includegraphics[width=.33\textwidth]{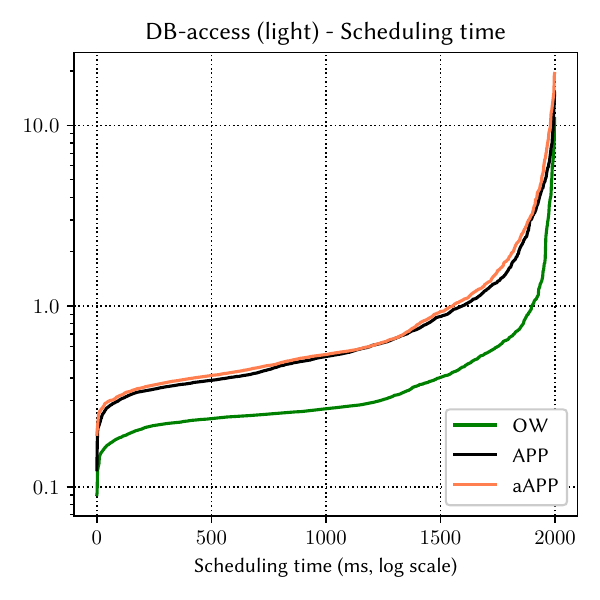}}
    \raisebox{-3em}{\includegraphics[width=.33\textwidth]{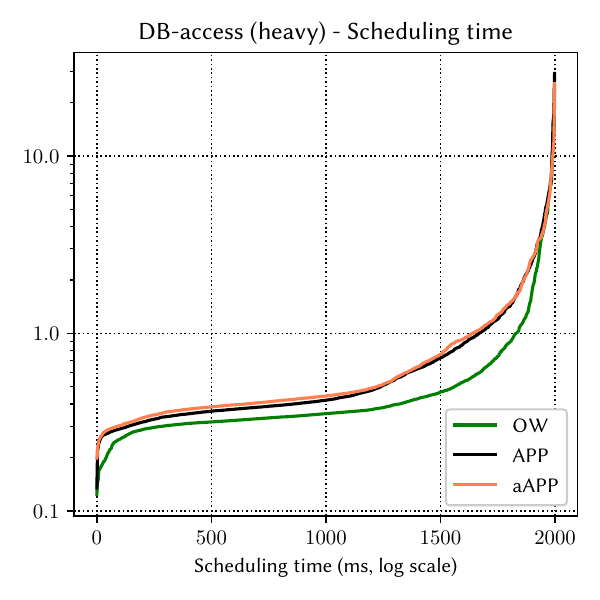}}
    \raisebox{-3em}{\includegraphics[width=.33\textwidth]{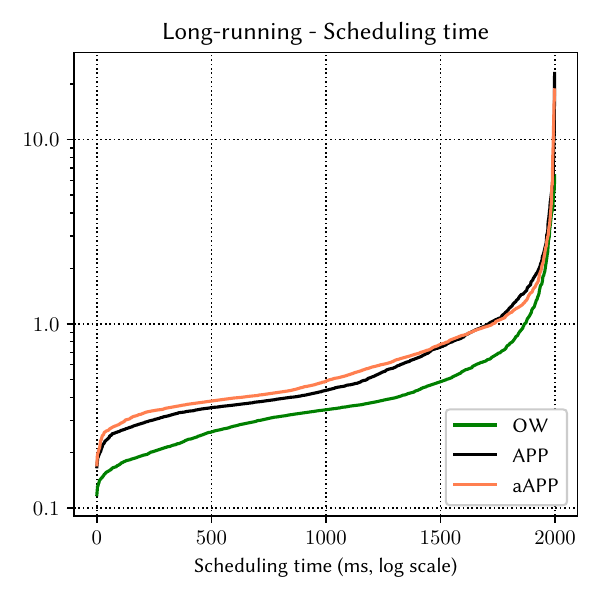}}
    \caption{Scheduling times distributions between OpenWisk, \app-, and \appp. }
    \label{fig:all_sched_overhead}
\end{figure}

\begin{figure}[h]
    \raisebox{-3em}{\includegraphics[width=.33\textwidth]{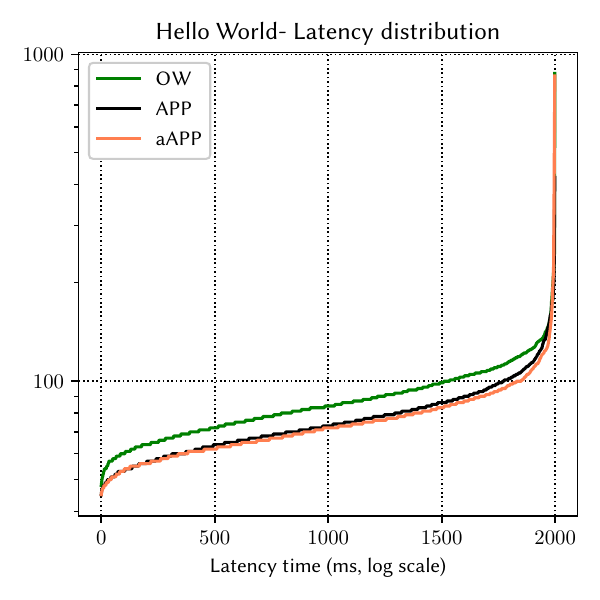}}
    \raisebox{-3em}{\includegraphics[width=.33\textwidth]{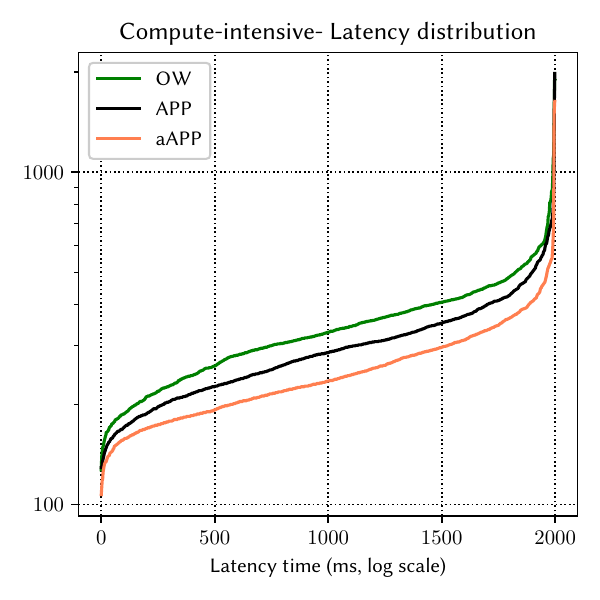}}
    \raisebox{-3em}{\includegraphics[width=.33\textwidth]{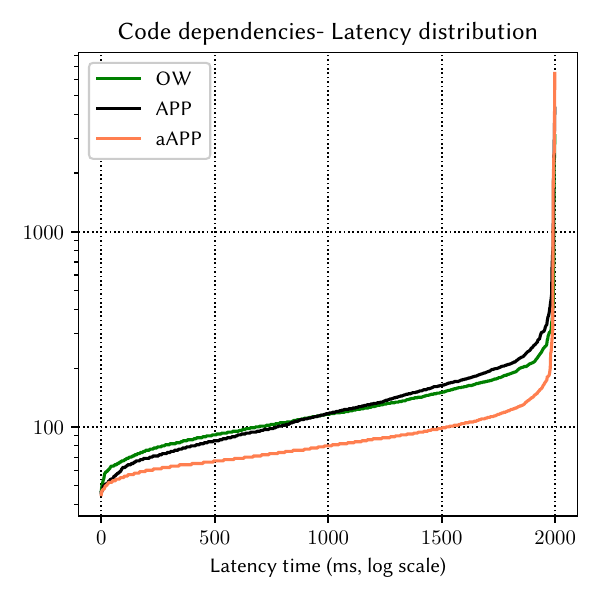}}
    \raisebox{-3em}{\includegraphics[width=.33\textwidth]{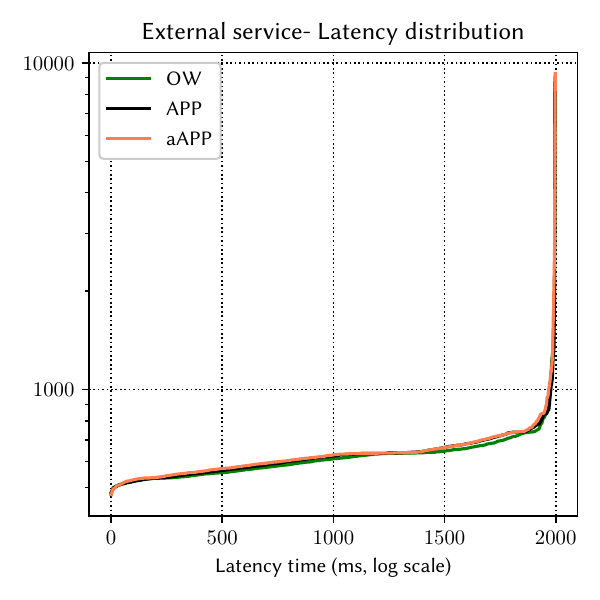}}
    \raisebox{-3em}{\includegraphics[width=.33\textwidth]{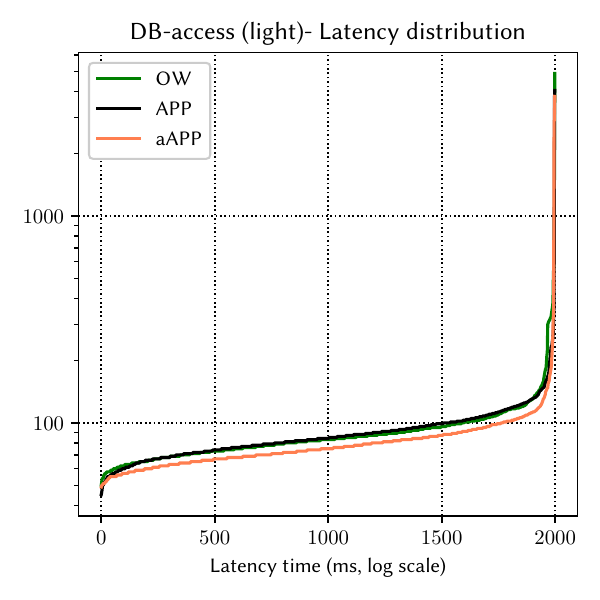}}
    \raisebox{-3em}{\includegraphics[width=.33\textwidth]{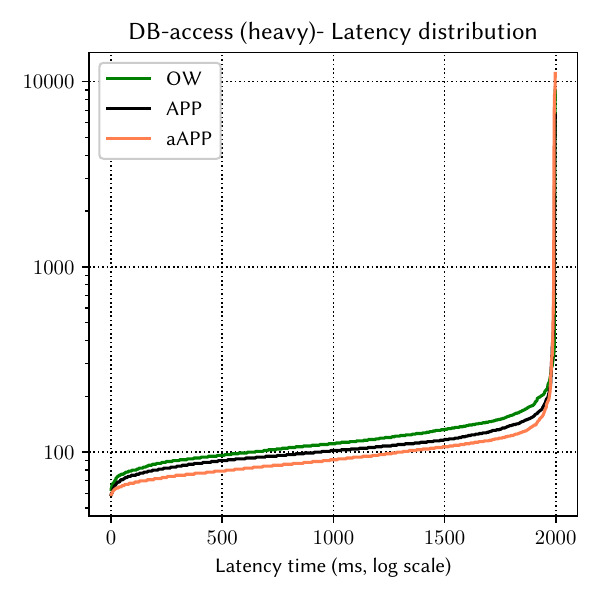}}
    \raisebox{-3em}{\includegraphics[width=.33\textwidth]{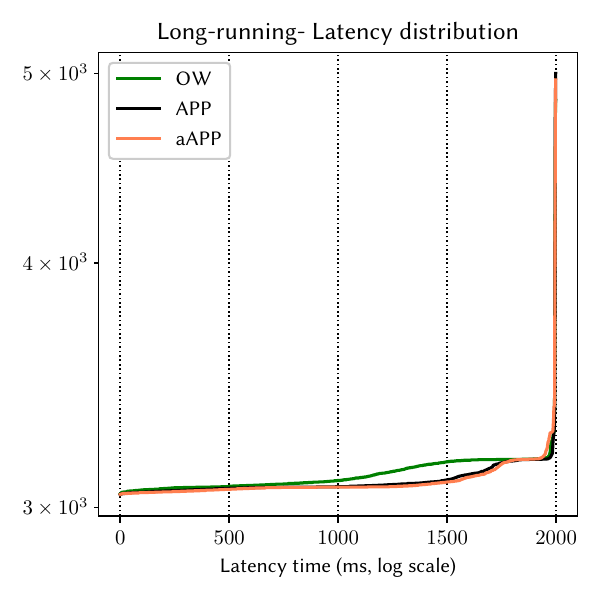}}
    \caption{Latency times distributions between OpenWisk, \app-, and \appp.}
    \label{fig:all_latency_overhead}
\end{figure}

We also show the plot-line distribution of the scheduling times of all
cases---whose average and standard deviations are in the submitted paper, along
with two representative plots---in \cref{fig:all_sched_overhead}; in
\cref{fig:all_latency_overhead} we present the plot-line distribution of the
execution latencies.
%

We comment on the distributions, which complement the aggregate data from
\cref{tab:latencies}.

The results in \cref{fig:all_sched_overhead} follow our expectations. OpenWhisk
vanilla slightly outperforms APP, which in turn slightly outperforms aAPP. As
the platforms incrementally add new, more refined features, it seems reasonable
to have a small scheduling overhead when taking into account the new operations
performed by the load balancer.

Interestingly, if we consider the latency of the results (in
\cref{fig:all_latency_overhead}), it appears that \appp slightly outperforms
OpenWhisk. We ascribe this behaviour to the high variability (as per the
standard deviation in \cref{tab:latencies}) of performance of the cloud
instances and the inherent variability of the cases, as witnessed also by the
considerable curve tails. 



\bibliography{biblio}